\documentclass[12pt]{article}
 \pdfoutput=1
\usepackage{amsfonts,amsthm,amsmath,amssymb,slashed}
\usepackage[textwidth = 500 pt, textheight = 630 pt]{geometry}
\usepackage{latexsym,amsfonts,amsmath,amssymb,theorem,dsfont,wasysym}
\usepackage{amssymb}
\usepackage{amsmath}
\usepackage{amstext}
\usepackage{graphicx,epsfig}
\usepackage{epsfig}
\usepackage{verbatim} 
\usepackage{fancyhdr}
\usepackage{fancybox}
\usepackage{color}
\usepackage{ulem,bbold}
\usepackage{enumitem}
\usepackage{subfigure}
\usepackage{bbm}
\usepackage{parskip}
\usepackage[numbers,sort&compress]{natbib}

\definecolor{MyDarkBlue}{rgb}{0.15,0.15,0.45}
\usepackage[linktocpage=true]{hyperref}
\hypersetup{
colorlinks=true,
citecolor=MyDarkBlue,
linkcolor=MyDarkBlue,
urlcolor=MyDarkBlue,
}

\usepackage[numbers,sort&compress]{natbib}

\def\beq{\begin{eqnarray}}
\def\eeq{\end{eqnarray}}

\def\({\left(}
\def\){\right)}

\def\FI{\hat{\phi}_I}

\newcommand{\be}{\begin{equation}}
\newcommand{\ee}{\end{equation}}
\newcommand{\la}{\langle}
\newcommand{\ra}{\rangle}
\def\ea{\end{eqnarray}}
\def\ba{\begin{eqnarray}}

\def\beq{\begin{eqnarray}}
\def\eeq{\end{eqnarray}}

\def\({\left(}
\def\){\right)}

\def\la{\langle}
\def\ra{\rangle}

\def\lsim{\mathrel{\rlap{\lower3pt\hbox{\hskip0pt$\sim$}}
     \raise1pt\hbox{$<$}}}         
\def\gsim{\mathrel{\rlap{\lower4pt\hbox{\hskip1pt$\sim$}}
     \raise1pt\hbox{$>$}}}         

\def\lsim{\mathrel{\rlap{\lower3pt\hbox{\hskip0pt$\sim$}}
     \raise1pt\hbox{$<$}}}         
\def\gsim{\mathrel{\rlap{\lower4pt\hbox{\hskip1pt$\sim$}}
     \raise1pt\hbox{$>$}}}         

\usepackage{amsmath}
\usepackage{amsfonts}
\usepackage{verbatim}
\usepackage{graphicx}
\usepackage{subfigure}
\usepackage{amssymb}
\usepackage[T1]{fontenc}
\usepackage{slashed}

\begin{document}

\renewcommand{\thefootnote}{\fnsymbol{footnote}}

\makeatletter
\@addtoreset{equation}{section}
\makeatother
\renewcommand{\theequation}{\thesection.\arabic{equation}}

\rightline{}
\rightline{}



\begin{center}

{\Large \bf{On Evolution of Coherent States\\ as Quantum Counterpart of Classical Dynamics}}

 \vspace{1truecm}
\thispagestyle{empty} \centerline{\large  {Lasha  Berezhiani and Michael Zantedeschi}
}

\textit{Max-Planck-Institut f\"ur Physik, \\ F\"ohringer Ring 6, 80805 M\"unchen, Germany\\}
\vskip 10pt
\textit{Arnold Sommerfeld Center, Ludwig-Maximilians-Universit\"at,\\ Theresienstra{\ss}e 37, 80333 M\"unchen, Germany}

\end{center}  
 
\begin{abstract}

Quantum dynamics of coherent states is studied within quantum field theory using two complementary methods: by organizing the evolution as a Taylor series in elapsed time and by perturbative expansion in coupling within the interaction-picture formalism. One of the important aspects of our analysis consists in utilizing the operators and the vacuum of interacting theory in constructing the states, without invoking asymptotic particles. Focusing on a coherent state describing a spatially homogeneous field configuration, it is demonstrated that both adopted methods successfully account for nonlinear classical dynamics, giving distinguishable quantum effects. In particular, according to the time-expansion analysis the initial field-acceleration, with which the field departs from its initial expectation value, is governed by the tree-level potential with renormalized mass and bare coupling constant. The interaction-picture computation, instead, can be manipulated to give the nonlinear dynamics, determined in terms of renormalized coupling and mass. However, it results in a logarithmic initial-time singularity in the field-acceleration, reminiscent of the similar behaviour encountered within semi-classical formalism, for certain choices of the initial state for fluctuations. Within our coherent-state analysis, the above mentioned peculiarities are artefacts of an expansion: in the first case over infinitesimal time, while in the second case in the coupling constant.
Despite this, we show that the evolution obtained within the interaction-picture analysis is valid for extended period of time. Moreover, on top of the desired classical dynamics, it serves us with interesting quantum corrections, previously proposed by Dvali--Gomez--Zell.

\end{abstract}

\newpage
\setcounter{page}{1}

\renewcommand{\thefootnote}{\arabic{footnote}}
\setcounter{footnote}{0}

\linespread{1.1}
\parskip 4pt

\section{Introduction}

\label{introduction}

As it is well known, the classical field configurations merely serve as an approximate description of the system. Within the framework of quantum field theory, the ultimate description of the state is given by a quantum state, which is an element of the physical Hilbert space. Although the classical description is extremely accurate for macroscopic systems, there could be still some cumulative quantum effects that could become relevant in time. In \cite{Dvali:2011aa,Dvali:2012en,Dvali:2012wq,Dvali:2013vxa,Dvali:2013eja,Dvali:2017eba}, the quantum corpuscular approach to classical backgrounds was developed and it was shown that the coherent state description of time-dependent systems leads to new type of quantum effects, which, in certain cases, can result in complete breakdown of the classical description. This was shown to have potentially important ramifications for systems such as black holes, de Sitter spacetime and cosmic inflation, predicting new deviations from semi-classical evolution. The importance of these considerations in the context of the beginning of inflation was discussed in \cite{Berezhiani:2015ola}\footnote{See also \cite{Berezhiani:2016grw}, where the origin of the observed density perturbations was identified, within the proposal of \cite{Dvali:2013eja}, as the quantum uncertainty in the number of inflaton constituents (with significantly screened mass-gap, due to collective gravitational effects) within the Hubble patch.}.

To be specific, let us begin by recapping the argument of \cite{Dvali:2017eba}. In a weakly interacting theory possessing a dimensionless quantum coupling $\alpha\ll 1$, one may consider a system in a quantum state with large occupation number $N$. Initially, the dynamics of such a system should be well approximated by its classical equations of motion. Usually, one may define a classical collective coupling $\alpha N$, which characterizes the strength of classical nonlinearities in the equation of motion and consequently sets the classical time-scale $t_{\rm cl}$, after which nonlinear corrections to the classical dynamics become important, even if initially the system was prepared in a state well-described by free waves. Quantum mechanically, the latter corresponds to the expectation value of the quantum field in a coherent state, with the mean occupation number determined by the amplitude of the wave in question. It was argued in \cite{Dvali:2017eba} that the quantum scattering among the constituent quanta should lead to the decoherence of the coherent state, which in turn might cause a quantum departure from the classical evolution, with a time-scale of significant deviation estimated as
\beq
t_{\rm q}=\frac{t_{\rm cl}}{\alpha}\,;
\label{introtq}
\eeq
referred to as \textit{quantum break-time} in \cite{Dvali:2017eba}\footnote{However, it was shown in \cite{Dvali:2013vxa} that the quantum break-time is in general shorter for systems near criticality (i.e. $\alpha N=1$) and possessing semi-classical Lyapunov exponent.}. There, these concepts were shown to take a particularly simple form, when applied to the massive scalar field with quartic self-interaction. For this system, a harmonically oscillating homogeneous field represents an accurate solution to the full nonlinear equation on timescales
\beq
t\ll t_{\rm cl}\equiv\frac{1}{\omega (\alpha N)}\,;
\label{introtcl}
\eeq
where, according to the notation of \cite{Dvali:2017eba}, $\omega$ is the classical frequency of harmonic oscillations, the quantum coupling $\alpha\equiv \hbar \lambda$ is connected to the classical parameter $\lambda$ of quartic nonlinearity (with $\alpha\rightarrow 0$ in the classical limit of vanishing Planck's constant $\hbar\rightarrow 0$) and the occupation number (within the Compton volume $\sim \omega^{-3}$)  $N=\frac{A^2}{\hbar \omega^2}$ is determined in terms of the amplitude $A$ of the classical wave; implying the independence of $t_{\rm cl}$ from $\hbar$. Representing the classical background as a coherent state of zero-momentum particles and estimating the rate of quantum-breaking using  $2\rightarrow 2$ scattering of constituent quanta, it was shown in \cite{Dvali:2017eba} that the quantum break-time reduces to the general expression \eqref{introtq}. Although $t_{\rm q}\gg t_{\rm cl}$, it was argued in \cite{Dvali:2017eba} that the gradual quantum decoherence is a cumulative effect and should persist even after one properly accounts for the classical nonlinearities, extending the regime of validity beyond \eqref{introtcl}. Similar results were reproduced for axions in \cite{Dvali:2017ruz}. Recently, analogous conclusions on quantum breaking were drawn in \cite{Kovtun:2020ndc,Kovtun:2020kcl} , based on numerical analysis within the so-called \textit{2-Particle-Irreducible} (2PI) formalism for $(1+1)$-dimensional self-interacting scalar field theory with a conserved charge.

The adoption of these arguments to certain physical systems has staggering ramifications, as we have mentioned at the beginning of this section. Focusing on cosmic inflation, for concreteness, we reiterate the proposal of \cite{Dvali:2013eja} and its consequences for the paradigm.  Let us further restrict the discussion to the so-called ``$m^2\phi^2$-model'', for simplicity. At the end of the slow-roll phase, the scalar field begins to oscillate and behaves as a non-relativistic condensate. A quantum state corresponding to such a configuration consists of large number of zero-momentum $\phi$-particles. The gravitational background sourced by such a quantum condensate was argued to represent some sort of a condensate as well, constructed around the Minkowski vacuum by gravitational degrees of freedom. The motivation for the latter is obvious, if one ware to drain the scalar condensate the gravitational configuration would approach the Minkowski space. Rewinding time to the slow-roll stage of inflation, the similar description was suggested to hold. Although, due to significant change in the classical dynamics, the degrees of freedom making up the condensates can go significantly off-shell. In \cite{Dvali:2013eja}, the properties of the gravitational condensate, i.e. the dispersion relation of constituents, was determined by demanding the reproduction of the semi-classical properties of the quasi-de Sitter space. Quantum processes such as particle production during inflation was identified as the scattering among the condensate constituents. For example, the microscopic process behind the production of gravitational waves was identified as the annihilation of the quanta making up the gravitational background. This led to the conclusion that if these processes last long enough the quantum depletion of the background should become significant and invalidate the semiclassical description. In fact, it was established that if inflaton starts out in the regime of slow-roll eternal inflation (self-reproduction), the dynamics should become fully quantum before the scalar field reaches the bottom of the potential and ends inflation.

The goal of this work is to start taking first steps towards establishing the link between the S-matrix estimates of \cite{Dvali:2013eja} and the rigorous real-time dynamics of coherent states for setups such as cosmic inflation. As it will become clear shortly, the scope of the current work is limited and is by no means directly applicable to cosmological frameworks. However, our analysis provides an important foundation for more sophisticated analysis by pinpointing the possible advantages of designing the states in its entirety. 

In general, approaches to computing the corrections to the classical evolution fall in two main categories. The most commonly used one is the semiclassical analysis, in which one studies the evolution of perturbations around the classical background and evaluates the semiclassical back-reaction on the latter. The expansion, within these methods, can be organized w.r.t. different parameters (e.g. coupling, $\hbar$ and many others) and a plethora of different effects can be captured (for instance, 2PI approach capturing thermalization). See \cite{Cooper:1994hr,Habib:1995ee,Cooper:1996ii,Aarts:2000wi,Mihaila:2000sr,Berges:2001fi,Aarts:2002dj,berges,Borsanyi:2008ar} and references therein for interesting work on the subject.  A useful overview of such methods, along with a complete list of relevant references can be found in \cite{calzetta}. The goal of this article is not to utilize this methods per se, but instead we will be questioning some of their underlying assumptions, as it will become clear.
 The other category consists of replacing the classical background by a coherent state (since coherent states represent a proper quantum counterpart of classical field configurations) of large number of quanta and studying the evolution as a multi-particle scattering process \cite{Dvali:2011aa,Dvali:2012en,Dvali:2012wq,Dvali:2013vxa,Dvali:2013eja,Dvali:2017eba}. In cases when classical evolution is not significantly affected by interactions, the said state is made of on-shell (asymptotic) particles, as discussed above\footnote{See \cite{Glauber:1963tx,Kibble:1965zza} for the earlier work on representing classical configurations as coherent states.}. In the opposite situation, the latter approach requires invoking off-shell degrees of freedom \cite{Dvali:2013eja,Dvali:2017eba}, with the off-shell-ness connected with the properties (e.g. dispersion relation) of the constituent quanta being related to the classical background and  continuously changing in case of nontrivial classical evolution.
 
The approach adopted in this work belongs to the second category. The distinguishing feature is the construction of coherent states out of the vacuum of the interacting theory, canonical field-operator and its conjugate momentum, without invoking asymptotic approximation for them\footnote{Our construction falls within the category of generalized coherent states; for the overview of various aspects see \cite{Zhang:1990fy}.}. Our work is, in spirit, very similar to the study of \cite{Vachaspati}, where the quantum mechanical analog case was analysed. Within our method, the initial construction is exact and valid to all orders in perturbation theory. Eventually, we are forced to perform a loop expansion for practical reasons. We show that for the leading order quantum effects it would have sufficed to construct the coherent state in terms of asymptotic particles, but our improved construction still helps with some intermediate technical steps. Furthermore, the adopted construction ensures the consistency of the perturbative expansion for the state and the dynamics; e.g. for 2-loop calculation the coupling dependence of the state appears to be relevant. Moreover, we perform a rigorous computation of nonlinear contributions to the classical dynamics and its quantum entourage, reproducing the Dvali--Gomez--Zell formula \eqref{introtq} for the quantum break-time.

In this work, we therefore study the quantum evolution of a coherent state in an interacting quantum field theory. The analysis is performed for a scalar field with quartic self-interaction, but can be straightforwardly extended to other theories (to be presented elsewhere). In particular, we study the Lagrangian\footnote{From this point forward, we will be working in units $\hbar=c=1$.}
\beq
\mathcal{L}=-\Lambda-\frac{1}{2}Z(\partial_\mu \hat{\phi})^2-\frac{1}{2}Zm^2\hat{\phi}^2-\frac{\lambda}{4!}Z^2\hat{\phi}^4\,;
\label{introlag}
\eeq
where $\Lambda$, $Z$, $m$ and $\lambda$ are the bare vacuum energy, field normalization, mass  and coupling constant respectively; with all of them expected to be infinite, as usual (keeping in mind that $Z$ does not get a correction at 1-loop, whithin the theory at hand). The parameters are considered to be such that there is a single non-degenerate vacuum with vanishing field-value.

Just like in textbook calculations, we assume that there exists a vacuum state $|\Omega\ra$ with unit norm, which is the Hamiltonian eigenstate with lowest possible eigenvalue. Then, the coherent state corresponding to a classical field configuration with $\phi_{cl}(x)$ and $\pi_{cl}(x)$ can be constructed as
\beq
|C\ra=e^{-i\int d^3 x \left( \phi_{cl}(x)\hat{\pi}(x)-\pi_{cl}(x)\hat{\phi}(x) \right)}|\Omega\ra\,,
\eeq
with $\pi$ standing for the conjugate momentum. The convenient property of this state is that it has unit norm and satisfies the following to all orders in $\lambda$ at the initial moment of time $t=0$
\beq
&&\la C | \hat{\phi} |C\ra (t=0)=\phi_{cl}(x)\,,\\
&&\la C | \hat{\pi} |C\ra (t=0)=\pi_{cl}(x)\,.
\eeq
We adopt two complementary methods for studying the dynamics: 

\begin{itemize}

\item[ {\bf (I)}] Instead of applying standard field-theoretical methods, we propose a new approach based on the evaluation of physical quantities in a Taylor series in time $t$, elapsed since the initial moment. We calculate first several terms of the expansion, giving it initially in the form that is valid to all orders in coupling constant. Examining the coherent-state-expectation-value of the field operator, we find that all manifestly finite contributions are identical to the iterative solution to the classical equation of motion, at each calculated order in $t$. Remarkably, these are supplemented with terms involving vacuum expectation values of singular operators, some of which can be absorbed via the renormalization of parameters, while the others need to be resummed together with higher order terms in $t$. We also found that the initial acceleration of this 1-point-function does not seem to be governed by the effective potential (a la Coleman-Weinberg \cite{Coleman:1973jx}), contradicting the standard intuition. Presumably, this too needs to be resolved through the resummation.
On the flip side, one advantage of the method in question is the absence of the so-called initial-time singularity, one tends to encounter within semi-classical techniques for certain initial states for fluctuations; for the relevant discussion of the latter and the proposed resolutions see \cite{Cooper:1987pt,Baacke:1997zz,Boyanovsky:1998aa,Baacke:1999ia,calzetta}.

\item[{\bf (II)}] In the attempt to extend the analysis to finite time-intervals and to ameliorate the above-mentioned puzzle with leftover divergencies, we have reanalyzed the evolution of the one-point expectation value in a coherent state using the interaction-picture approach, by performing a coupling expansion from the get-go. It is demonstrated that the expectation value of the field-operator evolves according to the classical equations of motion, albeit with renormalized parameters, supplemented with finite (yet small) quantum corrections. The price for this elegant result is the logarithmic initial-time singularity in the second-time-derivative of the one-point function in question, along with usual spurious secular divergencies that seem to be under control for an extended period of time. As we will show, expanding the result before the final step of the calculation in a Taylor series in elapsed time, we recover the results of the previous method. Most importantly, we will discuss the caveats concerning this reproduction. Although these results are limited by a given order of classical nonlinearities, it should be straightforward to extend them to higher orders.

\end{itemize}
\vskip 5pt

Furthermore, we use method (I) to evolve the state itself and compare it to the classically evolved coherent state, finding the departure after infinitesimal interval of time with interesting contributions. In (3+1)D, these corrections appear to be divergent, without any obvious regularization in sight that could give a physically sound result. We are therefore left with the only logical conclusion that these infinities need to be resummed with higher-order corrections in infinitesimal time to give a finite final answer. Although parametrically the corrections at hand appear to be of interesting form, assertive claims do not seem possible at this point. However, we show that the results become finite in (1+1)D, providing us with a tractable departure from the classical evolution.

The goal of this article, is therefore the following: to build explicitly, at an operatorial level,  the simplest possible coherent state mimicking a classical configuration and to analyze its dynamics. Our approach is orthogonal to the commonly used background field methods (e.g. \cite{Baacke:1997zz}), where initial conditions for 1-point and 2-point correlation functions are specified based on semi-classical reasoning and by requiring the absence of the initial time singularity. Our work helps to elucidate the points that are obscured within these methods. In fact, our analysis demonstrates that if (non-squeezed, i.e. simplest) coherent states constructed consistently, without even invoking asymptotic degrees of freedom and the free vacuum, are to be physical then the so-called "initial-time singularity" must be an artefact of the perturbative expansion. Or, one would have to proclaim the inconsistency of non-squeezed coherent states in four-dimensional quantum field theory in question.

The article is organized as follows. In Sec. 2, the theoretical framework is described. In Sec. 3, the main results of the time expansion are shown for a concrete coherent state, corresponding to the homogeneous classical field configuration. In particular we evaluate both the overlap between the quantum and classically evolved coherent states, and the expectation value of some operators. In Sec. 4, we perform the interaction-picture computation of the one-point expectation value in the coherent state. Finally, we summarize the results in Sec. 5 and discuss the outlook. Some of the technical details have been relegated to appendices.

\section{Setup}
\label{setup}

The Hamiltonian corresponding to the Lagrangian density \eqref{introlag} is given by
\beq
\hat{H}=\int d^3x\left( \frac{1}{2}\frac{\hat{\pi}^2}{Z}+\frac{1}{2}Z(\partial_j \hat{\phi})^2+\frac{1}{2}Zm^2\hat{\phi}^2+\frac{\lambda}{4!}Z^2\hat{\phi}^4 +\Lambda\right)\,,
\label{hamilton}
\eeq
where the conjugate momentum is given by $\hat{\pi}=Z\partial_t\hat{\phi}$. We also have the usual equal-time canonical commutation relations
\beq
&&[\hat{\phi}(x,t),\hat{\pi}(y,t)]=i\delta^{(3)}(x-y)\,,\\
&&[\hat{\phi}(x,t),\hat{\phi}(y,t)]=[\hat{\pi}(x,t),\hat{\pi}(y,t)]=0\,.
\eeq

Let us begin by constraining the parameters of the theory so that the energy density of the interacting vacuum $|\Omega\ra$ vanishes, i.e. by requiring
\beq
\label{zeroenergylevel}
\la \Omega | \hat{\mathcal{H}} |\Omega\ra=0\,,
\eeq
with $\mathcal{H}$ denoting the Hamiltonian density operator. For \eqref{hamilton} we get
\beq
\la \Omega | \hat{\mathcal{H}} |\Omega\ra=\Lambda+\frac{1}{2Z}\la \Omega | \hat{\pi}^2  |\Omega\ra+\frac{1}{2}Z\la \Omega | (\partial_j \hat{\phi})^2  |\Omega\ra+\frac{1}{2}Zm^2 \la \Omega | \hat{\phi}^2  |\Omega\ra+\frac{\lambda}{4!}Z^2\la \Omega | \hat{\phi}^4  |\Omega\ra\,.
\eeq
This can be made to vanish by adjusting the bare vacuum energy $\Lambda$ accordingly.

As already mentioned in the introduction, the coherent states of our main interest can be parameterised as
\beq
|C\ra=e^{-i\hat{f}}|\Omega\ra\,, \quad \text{with} \quad \hat{f}\equiv \int d^3 x \left( \phi_{cl}(x)\hat{\pi}(x,0)-\pi_{cl}(x)\hat{\phi}(x,0) \right)\,.
\label{cohst}
\eeq
As the expression speaks for itself, we have presented the state in the Heisenberg picture; moreover, notice that due to the definition of the conjugate momentum we have $\pi_{cl}=Z\dot{\phi}_{cl}$.

The Hamiltonian density \eqref{hamilton} in a coherent state \eqref{cohst} can be readily obtained, after adjusting the vacuum energy to zero, reducing to
\beq
\la C| \hat{\mathcal{H}}(x) |C \ra(t=0)=Z\left[\frac{1}{2}\frac{\pi_{cl}^2}{Z^2}+\frac{1}{2}\left(\vec{\nabla}\phi_{cl}\right)^2+\frac{1}{2}\left(m^2+\frac{\lambda Z}{2}\la\Omega|\hat{\phi}(x)^2 |\Omega\ra\right)\phi_{cl}^2+\frac{\lambda}{4!}Z\phi_{cl}^4\right]\,.
\label{initiclassham}
\eeq
Here and throughout this work, we employ the fact that $\la\Omega| \hat{\phi}|\Omega \ra=\la\Omega| \hat{\phi}^3|\Omega \ra=\la\Omega| \hat{\pi}|\Omega \ra=0\,,$ due to $Z_2$ symmetry.

This looks encouragingly similar to the classical expression
\beq
\mathcal{H}_{cl}=\frac{1}{2}\pi_{cl}^2+\frac{1}{2}\left(\vec{\nabla}\phi_{cl}\right)^2+\frac{1}{2}m_{\rm ph}^2\phi_{cl}^2+\frac{\lambda_{\rm ph}}{4!}\phi_{cl}^4\,.
\eeq
However, there are important differences as well. First thing to notice is the appearance of the divergent equal-point 2-point function. At this point, one might be inclined to absorb this divergence into the mass. In other words, at 1-loop order (at which, we expect $Z=1$) one could define a physical mass and coupling as
\beq
\label{massrenorm1}
&&m_{\rm ph}^2\stackrel{?}{\equiv} \left(m^2+\frac{\lambda}{2}\la\Omega|\hat{\phi}(x)^2 |\Omega\ra\right)\,,\\
&&\lambda_{\rm ph}\stackrel{?}{\equiv}\lambda \,;
\label{eq:renormalization_vacuum}
\eeq
which follows from identifying $\mathcal{H}_{cl}$ with the expectation value of $\hat{\mathcal{H}}$. Although the mass renormalization \eqref{massrenorm1} is of the expected form, the coupling assignment \eqref{eq:renormalization_vacuum} seems to be inconsistent with the standard one for our setup; but we will come to this point later on.

Before specializing on a coherent state describing a particular configuration of interest, let us make some general remarks that will aid us with technical calculations.

\subsection{Heisenberg Picture}

\label{heisenbergpicture}

Among other quantities, one might be interested in the expectation value of various operators in a coherent state. In general, it is convenient to discuss these in Heisenberg picture. Let us consider the following expectation value
\beq
\la C|\prod _i \mathcal{O}_i(\hat{\phi},\hat{\pi};t_i) | C \ra\,,
\label{expval}
\eeq
with $|C\ra$ being a coherent state \eqref{cohst} constructed using operators at time $t=0$ and $\mathcal{O}_i$ standing for a composite operator at a moment $t_i$; in other words, we have equal time products within each $\mathcal{O}$. Then it is straightforward to show that \eqref{expval} can be expressed as  a vacuum-expectation-value of shifted operators $\mathcal{O}_i(\phi_{cl}+\hat{\phi},\pi_{cl}+\hat{\pi};t_i)$ evolved from $t=0$ using the background-field-method Hamiltonian $\hat{H}[\phi_{cl}+\hat{\phi},\pi_{cl}+\hat{\pi}]$. Namely, we have
\beq
\la C|\ \prod_i\mathcal{O}_i(\hat{\phi},\hat{\pi};t_i) | C \ra=\la \Omega|\prod_i \left(e^{i\hat{H}[\phi_{cl}+\hat{\phi},\pi_{cl}+\hat{\pi}]t_i} \mathcal{O}_i(\phi_{cl}+\hat{\phi},\pi_{cl}+\hat{\pi};0)e^{-i\hat{H}[\phi_{cl}+\hat{\phi},\pi_{cl}+\hat{\pi}]t_i}\right) | \Omega \ra\,,~~
\label{heisen}
\eeq
which is quite intuitive as the exponential operator defining our coherent state \eqref{cohst} is a field displacement operator. For the face value, it appears as if we have evaluated the expectation value in semi-classical approximation by quantising perturbations around a fixed classical background and treating perturbations to be in the vacuum. However, due to the fact that $|\Omega\ra$ is the vacuum of the fundamental theory rather than the one of a semi-classical theory of perturbations, the matters are somewhat more complicated. As we will see on concrete examples, while the dynamics of correlators is identical to the one inferred from the background field method, the initial conditions are quite different. Nevertheless, it is important to show that \eqref{heisen} is exact and holds for any theory. It follows from the following mathematical identity
\beq
e^{e^{\hat{X}}\hat{Y}e^{-\hat{X}}}=\sum_{n=0}^\infty \frac{1}{n!}(e^{\hat{X}}\hat{Y}e^{-\hat{X}})^n=e^{\hat{X}} e^{\hat{Y}}e^{-\hat{X}}\,,
\label{identity}
\eeq
which holds for any operators $\hat{X}$ and $\hat{Y}$; together with the observation that for equal time operators we have
\beq
e^{i\hat{f}}\mathcal{O}(\hat{\phi},\hat{\pi};0)e^{-i\hat{f}}=\mathcal{O}(\phi_{cl}+\hat{\phi},\pi_{cl}+\hat{\pi};0)\,, 
\eeq
following from the Baker--Campbell--Hausdorff formula.

Regarding the coherent states themselves, we are parameterising them in terms of two functions of spatial coordinates $\phi_{cl}(x)$ and $\pi_{cl}(x)$, and a time stamp appearing on the operators invoked in their construction. Using \eqref{identity}, we can write
\beq
|\phi_{cl},\pi_{cl};t\ra={\rm exp}\left[ -i\int d^3 x \left( \phi_{cl}(x)\hat{\pi}(x,t)-\pi_{cl}(x)\hat{\phi}(x,t) \right)\right]|\Omega\ra\nonumber \\
={\rm exp}\left[ -i\int d^3 x e^{i\hat{H}t}\left( \phi_{cl}(x)\hat{\pi}(x,0)-\pi_{cl}(x)\hat{\phi}(x,0) \right)e^{-i\hat{H}t}\right]|\Omega\ra\nonumber \\
=e^{i\hat{H}t}{\rm exp}\left[ -i\int d^3 x \left( \phi_{cl}(x)\hat{\pi}(x,0)-\pi_{cl}(x)\hat{\phi}(x,0) \right)\right]e^{-i\hat{H}t}|\Omega\ra\nonumber\\
=e^{i\hat{H}t}|\phi_{cl},\pi_{cl};0\ra\,.
\label{heisenstate}
\eeq
Notice that this relation is similar to the one we have for field (and conjugate momentum) eigenstates in Heisenberg picture. 

\subsection{Schr\"odinger Picture}
\label{schrodpic}

Besides expectation values, one is usually interested in transition amplitudes as well, for which it may be convenient to see how the coherent state itself is evolving.

To see how the coherent state \eqref{cohst} evolves in the Schr\"odinger picture, we use \eqref{identity} once more to obtain
\beq
|C\ra (t)=e^{-i\hat{H}t} e^{-i\hat{f}}|\Omega \ra=e^{-ie^{-i\hat{H}t}\hat{f}e^{i\hat{H}t}}e^{-iE_{\Omega}t}|\Omega \ra\,;
\label{ct}
\eeq
with $E_\Omega$ denoting the vacuum eigenvalue of the Hamiltonian, which we have been adjusting to zero \eqref{zeroenergylevel}.
In other words, the evolved state is the one created by the same linear combination of $\hat{\phi}$ and $\hat{\pi}$, but 'evolved' backwards in time. This seems to correspond to a nontrivial squeezing of the coherent state. Moreover, comparing to the Heisenberg picture result, time-evolution simply scans through states \eqref{heisenstate}, with fixed $\phi_{cl}(x)$ and $\pi_{cl}(x)$.

\section{Concrete Example: Homogeneous background}

A state of particular interest is the one corresponding to the homogeneous field configuration, as being a popular representative of configurations used in background field calculations. In particular, we focus on a state that corresponds to a classical homogeneous background with the following initial conditions
\beq
&&\phi_{cl}(t=0)=\phi_0\,,\\
&&\pi_{cl}(t=0)=0\,.
\eeq
Here we have chosen vanishing initial momentum for simplicity. Classically, this background will undergo anharmonic oscillations. In order to study the quantum aspects of this evolution, we need to define the quantum state corresponding to this classical background.

Following our previous discussion we have (for some rudimentary discussion see appendix A)
\beq
|C\ra(t=0)=e^{-i\hat{f}}|\Omega\ra\,,\qquad \text{with}\qquad \hat{f}\equiv \int d^3x\phi_0\hat{\pi}\,;
\label{initstate}
\eeq

In the remainder of this section we will compute the rate of departure of this state from the classical evolution and the expectation values of different operators.

\subsection{Rate of Departure}
\label{departurerate}

Applying \eqref{ct} to the homogeneous state at hand, we get
\beq
|C\ra (t)=e^{-i\phi_0\int d^3x e^{-i\hat{H}t}\hat{\pi}(x)e^{i\hat{H}t}}|\Omega \ra\,.
\label{ct1}
\eeq
Here we choose to work in Schr\"odinger picture, hence there is no time-label on $\pi$. At this point, we are interested in analysing the evolution of the state for an infinitesimal time $t$  and comparing the result to the classical evolution. In particular, we will be interested in evolving the state up to $t^2$-order, since this is where interesting effects begin to emerge. This is done, using
\beq
\int d^3x e^{-i\hat{H}t}\hat{\pi}(x)e^{i\hat{H}t}=\int d^3x\left[\hat{\pi}(x)+t\left(Zm^2\hat{\phi}+\frac{\lambda Z^2}{3!}\hat{\phi}^3 \right)-\frac{t^2}{2}\left( m^2\hat{\pi} +\frac{\lambda Z}{4}\left(\hat{\pi}\hat{\phi}^2+\hat{\phi}^2\hat{\pi}\right)\right)\right]\nonumber \\+\mathcal{O}(t^3)\,.
\eeq

Now, classically we know how the scalar field evolves in infinitesimal time $t$. Namely, if at $t=0$ we had $\phi=\phi_0$ and $\dot{\phi}=0$ then at time $t$ we would have
\beq
\phi_{cl}(t)=\phi_0-\frac{t^2}{2}\left( m_{\rm ph}^2 \phi_0+\frac{\lambda_{\rm ph}}{3!}\phi_0^3 \right)+\mathcal{O}(t^3)\,.
\eeq

In order to parameterise the deviation of an evolved coherent state from a coherent state constructed for the latter classical background, we calculate the overlap between \eqref{ct} and
\beq
|C_{cl}\ra (t)=e^{-i\int d^3x\left( \phi_{cl}(t)\hat{\pi}-\pi_{cl}(t)\hat{\phi} \right) }|\Omega\ra\,.
\label{cohclass}
\eeq
Here, we will assume $\phi_{cl}(t)$ and $\pi_{cl}(t)=Z\dot{\phi}_{cl}(t)$ to be given by the above mentioned classical solution, albeit with renormalized parameters.

By calculating number of commutators, the above expression can be written as
\beq
\la C_{cl}|C(t)\ra=\la \Omega | e^{\hat{Y}(t)} |\Omega \ra\,.
\eeq
In the limit of infinitesimal time we can straightforwardly evaluate $\hat{Y}(t)$ up to $t^2$ order, arriving at
\beq
\hat{Y}(t)=&&it \left[ \frac{1}{2}Z\phi_0^2\left(m_{\rm ph}^2-m^2+\frac{\lambda_{\rm ph}-\lambda Z}{3!}\phi_0^2 \right)V+\frac{1}{24}\lambda Z^2\phi_0^4V\right.\nonumber \\
&&~~~~+Z\phi_0\left(m_{\rm ph}^2-m^2+\frac{\lambda_{\rm ph}-\lambda Z}{3!}\phi_0^2\right)\int d^3 x\hat{\phi}(x)\nonumber\\
&&~~~~\left. -\phi_0\int d^3 x\left(\frac{\lambda Z^2\phi_0}{4}\hat{\phi}^2+\frac{\lambda Z^2}{3!} \hat{\phi}^3\right) \right]\nonumber\\
&-&it^2\left[ \frac{1}{2}\phi_0\left(m_{\rm ph}^2-m^2+\frac{\lambda_{\rm ph}-\lambda Z}{3!}\phi_0^2 \right)\int d^3 x\hat{\pi} \right.\nonumber \\
&&~~~~~\left. -\frac{1}{2}\phi_0 \int d^3 x \left(\frac{\lambda Z\phi_0}{4}(\hat{\pi}\hat{\phi}+\hat{\phi}\hat{\pi}) +\frac{\lambda Z}{4}(\hat{\pi}\phi^2+\hat{\phi}^2\hat{\pi})\right)\right]+\mathcal{O}(t^3)\,,
\label{Z}
\eeq
where $V$ denotes the spatial volume, i.e $V\equiv\int d^3 x$.
Notice that if quantum and classical evolutions were identical, then we would end up with $\la C_{cl}|C(t)\ra=1$; of course, after taking into account a proper relation between the physical and bare parameters. Notice that this calculation is different from the infinitesimal-time transition amplitude analysis, used in path-integral formalism; the latter will be briefly discussed in section \ref{pathintegral}.

Let us proceed to calculating the probability for an evolved state to coincide with the classically evolved state, resulting in
\beq
|\la C_{cl}|C(t)\ra |^2=1-t^2 D+\mathcal{O}(t^3)\,,
\label{probdep}
\eeq
with
\beq
D=&&\frac{(\lambda Z^2)^2\phi_0^4}{16}\int d^3x_1d^3x_2 \left[ \la \hat{\phi}(x_1)^2\hat{\phi}(x_2)^2\ra-\la \hat{\phi}(x_1)^2\ra\la\hat{\phi}(x_2)^2\ra \right]\nonumber \\
&&+\frac{(\lambda Z^2)^2\phi_0^2}{36}\int d^3x_1d^3x_2 \la \hat{\phi}(x_1)^3\hat{\phi}(x_2)^3\ra\nonumber \\
&&-\frac{\lambda Z^3}{3}\phi_0^2\left( m_{\rm ph}^2-m^2+\frac{\lambda_{\rm ph}-\lambda Z}{3!}\phi_0^2\right)\int d^3x_1d^3x_2\la \hat{\phi}(x_1)\hat{\phi}(x_2)^3\ra\nonumber\\
&&+Z^2\phi_0^2 \left( m_{\rm ph}^2-m^2+\frac{\lambda_{\rm ph}-\lambda Z}{3!}\phi_0^2\right)^2 \int d^3x_1d^3x_2\la \hat{\phi}(x_1)\hat{\phi}(x_2)\ra\,.
\eeq
We would like to stress that in this expression, $\la\ldots\ra$  stands for the vacuum expectation value.
Simplifying this expression to order $\lambda^2$, we get
\beq
D=&&Z^2\phi_0^2\left( m_{\rm ph}^2-m^2-\frac{\lambda Z}{2}\la \hat{\phi}^2\ra+\frac{\lambda_{\rm ph}-\lambda Z}{3!}\phi_0^2\right)^2\int d^3x_1d^3x_2\la \hat{\phi}(x_1)\hat{\phi}(x_2)\ra\nonumber\\
&&+\frac{(\lambda Z^2)^2\phi_0^4}{8}\int d^3x_1d^3x_2\la \hat{\phi}(x_1)\hat{\phi}(x_2)\ra^2\nonumber \\
&&+\frac{(\lambda Z^2)^2\phi_0^2}{4}\int d^3x_1d^3x_2\la \hat{\phi}(x_1)\hat{\phi}(x_2)\ra^3+\mathcal{O}(\lambda^3)\,,
\label{dd}
\eeq
where we have dropped terms that were obviously higher order than desired.
Provided that $m_{\rm ph}$ and $\lambda_{\rm ph}$ are the renormalized mass and coupling, there is no possibility for the first line of \eqref{dd} to give $\lambda^2$ contributions.

Therefore, the expression for $D$ simplifies to the last two lines of \eqref{dd}, keeping in mind that, at the order we are working, all the expectation values need to be evaluated within free theory. Now, these remaining terms look interesting. In particular, the second line of \eqref{dd} looks like a semi-classical term, as it goes as $(\lambda N)^2$ (with $N\propto \phi_0^2$ being number of quanta within the Compton volume), while the third line behaves as $\lambda^2 N$.

The problem is the evaluation of the integrals. If we compute them by Fourier transforming the 2-point function and then we exchange the order of momentum and position integrals, we end up with a divergent result. The IR divergence can be regulated by taking a finite volume, however the integrals are also UV divergent. At this point the only way around the latter is to proceed via analytic continuation. For instance, using the Fourier space expression for the propagator and exchanging the order of integration we readily obtain
\beq
\int d^3x_1d^3x_2\la \hat{\phi}(x_1)\hat{\phi}(x_2)\ra^2=\frac{V}{4(2\pi)^3}\int d^3k \frac{1}{k^2+m^2}\,.
\label{d1}
\eeq
To perform the integration we use dimensional regularisation. After analytically continuing the result back to 3d, we get
\beq
\int d^3x_1d^3x_2\la \hat{\phi}(x_1)\hat{\phi}(x_2)\ra^2=-\frac{V}{16 \pi}m\,.
\label{reg1}
\eeq
The problem with the result is the negative sign on the right-hand side of \eqref{reg1} and a negative contribution to $D$ seems to violate unitarity (notice that for the face value the left-hand side seems manifestly positive definite). Analogously, the integral appearing in the third line of \eqref{dd} can be processed in the Fourier space as
\beq
\int d^3x_1d^3x_2\la \hat{\phi}(x_1)\hat{\phi}(x_2)\ra^3=\frac{V}{8(2\pi)^6} \int d^3k_1 d^3k_2 \frac{1}{\sqrt{k_1^2+m^2}\sqrt{k_2^2+m^2}\sqrt{(\vec{k}_1+\vec{k}_2)^2+m^2}}\,.
\label{d2}
\eeq
Regretfully, we were unable to evaluate this integral further. Also, it must be noted that the second line of \eqref{dd} must be made finite independently from the third line, as they scale differently in $\phi_0$.

In the absence of a regularization scheme, that yields a physically sound finite result, it seems our perturbative expansion in time breaks down here and the resummation is warranted. In other words, the only conceivable resolutions to the puzzle at hand seems to be the resummation of this infinity together with higher order terms in $t$, in order to hopefully achieve a finite answer. 

However, as one can see, the fate of $D$ is dimension sensitive. In fact, in (1+1)-dimensions both integrals from \eqref{d1} and \eqref{d2} converge. The first can be easily computed analytically; the second one, on the other hand, requires numerical integration. Putting results together, the expression reduces to
\beq
|\la C_{cl}|C(t)\ra _{2D}|^2=1-t^2 \cdot \lambda^2\phi_0^2\cdot\frac{L}{64m}\cdot\left(\frac{\alpha}{2}+\phi_0^2 \right)+\mathcal{O}(\lambda^3,t^3)\,,
\label{p2}
\eeq
with $\alpha\approx 3$ and $L$ denoting the spatial system size.

An important point to notice is the lack of a linear term in this expression of departure of probability from unity, which would have been the rate of departure. Instead, we have the acceleration of departure. This  is connected to the property of infinitesimal transition amplitudes for coherent states we are about to discuss in the following section.

Before moving forward to the evolution of a one-point expectation value, where the treatment of infinities can be discussed in a clearer way, it is worth pausing to see the fate of these infinities  within path-integral formalism for coherent states. In particular, we will see that the divergencies similar to those in \eqref{dd} do not pose a problem there.

\subsection{Path-Integral}
\label{pathintegral}

Let us now discuss the path-integral formalism for coherent states. This topic has a long history going back to \cite{Klauder:1960kt,Schweber,Berezin:1971jf,Klauder:1979gi} (see also \cite{Zhang:1990fy} and references therein). Here, we simply overview some of the intermediate steps in the derivation.

The main motivation for this section is to underline the difference between the philosophy of section \ref{departurerate} and the commonly accepted approach to infinitesimal segments in path-integral. For the coherent states \eqref{cohst} the latter can be calculated as
\beq
\la\phi'_c,\pi'_c;t+dt|\phi_c,\pi_c;t\ra=\la \Omega | e^{i\int d^3 x \left( \phi'_c(x)\hat{\pi}(x)-\pi'_c(x)\hat{\phi}(x) \right)}\cdot e^{-i\hat{H}dt} \cdot e^{-i\int d^3 x \left( \phi_c(x)\hat{\pi}(x)-\pi_c(x)\hat{\phi}(x) \right)} | \Omega \ra\,,
\label{gensegment}
\eeq
with $dt$ being an infinitesimal time-interval, while $\{\phi_c(x),\pi_c(x)\}$ and $\{\phi'_c(x),\pi'_c(x)\}$ stand for c-number functions characterising the initial and final states respectively.

We can use the tricks of section \ref{heisenbergpicture}, to show that \eqref{gensegment} can be rewritten as
\beq
\la\phi'_c,\pi'_c;t+dt|\phi_c,\pi_c;t\ra&=&e^{\frac{i}{2}\int d^3 z(\delta\phi_c(z)\pi_c(z)-\phi_c(z)\delta\pi_c(z))}\times\nonumber\\
&&\times\la \Omega | e^{i\int d^3 x \left( \delta\phi_c(x)\hat{\pi}(x)-\delta\pi_c(x)\hat{\phi}(x) \right)}\cdot e^{-i\hat{H}[\phi_c+\hat{\phi},\pi_c+\hat{\pi}]dt} | \Omega \ra\,,\nonumber
\eeq
without loss of generality; here, we have introduced the notation $\delta\phi_c\equiv \phi'_c-\phi_c$ and $\delta\pi_c\equiv \pi'_c-\pi_c$. Notice that the Hamiltonian is a functional of shifted fields.
At this point, it has not been assumed that as $dt\rightarrow 0$, field variations should also vanish.

If we were to assume infinitesimal variation of the field and its conjugate momentum, then the expression would reduce to its familiar form. In other words, treating $\delta\phi_c$ and $\delta\pi_c$ to be as infinitesimal as $dt$ and working to linear order in these quantities, we could easily derive the following simplified transition amplitude
\beq
\la\phi'_c,\pi'_c;t+dt|\phi_c,\pi_c;t\ra \Big|_{infinitesimal}=e^{i\left(-\bar{H}dt+\frac{1}{2}\int d^3 x(\delta\phi_c \pi_c-\phi_c \delta\pi_c)\right)}\,;
\label{smallsegment}
\eeq
where the c-number Hamiltonian $\bar{H}$ is defined as
\beq
\bar{H}\equiv \int d^3 x\left(\frac{1}{2}\pi_c^2+\frac{1}{2}(\vec{\nabla}\phi_c)^2+\frac{1}{2}\left(m^2+\frac{\lambda}{2}\la\Omega|\hat{\phi}^2|\Omega\ra\right)\phi_c^2+\frac{\lambda}{4!}\phi_c^4\right)\,.
\eeq
Notice the appearance of the same mass renormalization as in previous sections. This should give us a pause, since for the standard path-integral construction, with field eigenstates, there are no renormalizable infinities emerging at this stage of the derivation.

However, this assumption of infinitesimal $\delta\phi_c$ and $\delta\pi_c$ may not be, strictly speaking, warranted for the purposes of path integral derivation, as for the latter we need intermediate path segments with arbitrary configurations. A short time interval does not entail that only paths with small variation contribute. In the standard path-integral formalism, this complication is avoided due to the orthonormality of the field-eigenstate basis. Plausibly, one can make the case that most of the contribution comes from more or less smooth trajectories, but it is an approximation nonetheless. We need to be extra cautious here, especially considering that we are after tiny cumulative effects. It must be stressed that the limit of smooth paths is precisely the approximation that is used in the literature to construct the path-integral for coherent states \cite{Klauder:1960kt,suzuki,Zhang:1999is}.

Let us appreciate the fact that for smooth trajectories the transition amplitude for an infinitesimal segment takes such an elegant form \eqref{smallsegment}. In the standard case, that is if we had dealt with field-eigenstates, there would have been only a field configuration labelling  states and the path-integral over conjugate momenta on the right-hand-side.

In our case, on the other hand, for an infinitesimal-time-transition-amplitude that assumes $\delta\phi_c$ and $\delta\pi_c$ to be $\mathcal{O}(dt)$ or smaller, we get a pure phase. This is precisely the reason why $\mathcal{O}(t)$ terms are lacking in \eqref{probdep}.

Going along with the approximation at hand, we can use \eqref{smallsegment} to construct the path-integral over a finite period of time by piling up a large number of infinitesimal segments and integrating over intermediate configurations, keeping in mind that coherent states provide us with a resolution of identity. In so many words, we can write
\beq
\la\phi'_{c},\pi'_{c};t'|\phi_{c},\pi_{c};t\ra \simeq \int [d\mu(\phi,\pi)] e^{i\int_{t}^{t'} dt\left(\int d^3x \dot{\phi}\pi-\bar{H}\right)}\cdot e^{\frac{i}{2}\int d^3x \left[\pi_c'\phi'_c-\pi_{c}\phi_{c}\right]}\,,
\label{pathintegral1}
\eeq
with the measure $[d\mu]$ defined using the following resolution of identity
\beq
1=\int [d\mu(\phi,\pi)]~ |\phi,\pi\ra\la \phi,\pi |\,,
\eeq
implying the measure being dimensionless; for the detailed discussion see \cite{Zhang:1990fy} and references therein (a brief discussion of Green's functions in this formalism is given in appendix B).

Alternatively, one could follow the standard path-integral construction, utilizing the resolution of identity by means of field and conjugate momentum eigenstates. In that case, the coherent state appears only in wave-functions of the initial and final states
\beq
\la \phi_c',\pi_c';t' |\phi_c,\pi_c;t \ra=\int\mathcal{D}\phi\,\mathcal{D}\pi~{\rm exp}\left({i\int_{t}^{t'} d^4x\left( \pi\dot{\phi}-\mathcal{H}(\phi,\pi) \right)}\right)\cdot\la\Omega|\phi-\phi_c';t'\ra\cdot\la\phi-\phi_c;t |\Omega\ra\nonumber\\
\cdot e^{-\frac{i}{2}\int d^3x\left( \phi_c\pi_c-\phi_c'\pi_c' \right)}\cdot e^{i\int d^3x~(\phi(t)\pi_c-\phi(t')\pi_c')}\,.~~~
\label{pathintegral2}
\eeq
Here,  $|\phi;t\ra$ appearing on the right-hand-side represents an eigenstate of $\hat{\phi}(t)$.

Let us finish this section by pointing out important differences between \eqref{pathintegral1} and \eqref{pathintegral2}. For starters, the Hamiltonians differ by the infinite contribution that should be absorbed in the renormalization of mass. Furthermore, we did not have to assume smooth paths in the derivation of \eqref{pathintegral2}, making it somewhat superior to \eqref{pathintegral1}. The price of this elegant result is the appearance of the wave-function of the vacuum, which would have to be calculated perturbatively. These wave-functions can be further simplified by shifting the field-configuration of integration by the classical function satisfying the appropriate boundary conditions. The last factor from \eqref{pathintegral2} simply imposes the initial and final conditions on the conjugate momentum. 

We will not discuss further the path-integral formalism in this work. It was mentioned purely for comparative purposes of its infinitesimal segment to the one computed in the previous subsection.

\subsection{Expectation Value of the Field}
\label{fieldexpval}

We would like to calculate the evolution of the expectation value of the field operator in the coherent state  \eqref{initstate}, that is $\la C| \hat{\phi} |C \ra (t)$, and compare it to its classical counterpart. Applying the general Heisenberg picture formalism we laid out in section \ref{heisenbergpicture} and expanding the result to order $t^4$, we arrive at the following expression\footnote{Some of the relevant commutators can be found in appendix C.}
\beq
\la C| \hat{\phi}(x) |C\ra (t)=\phi_0-\frac{t^2}{2}\left[ \phi_0\left(m^2+\frac{\lambda Z}{2}\la\Omega |\hat{\phi}(x)^2 |\Omega\ra\right)+\frac{\lambda Z}{3!}\phi_0^3\right]-\frac{t^3}{3!}\left[\frac{\lambda}{2}\phi_0\la \Omega |\hat{\pi}\hat{\phi}+\hat{\phi}\hat{\pi} | \Omega\ra\right]\nonumber\\
+\frac{t^4}{4!}\left[ \lambda Z\phi_0\la \Omega | (\partial_\mu \hat{\phi})^2+m^2\hat{\phi}^2 |\Omega\ra+
\phi_0\left( m^4+\lambda Zm^2\la\Omega| \hat{\phi}^2|\Omega\ra+\frac{5}{2\cdot 3!}(\lambda Z)^2\la \Omega | \hat{\phi}^4 |\Omega\ra \right)\right. \nonumber \\
 \left. +\frac{\lambda Z}{3!}\phi_0^3\left( 4m^2+5\lambda Z\la\Omega | \hat{\phi}^2|\Omega\ra \right)+\frac{(\lambda Z)^2}{2\cdot 3!}\phi_0^5\right]+\mathcal{O}(t^5)\,.~~~
 \label{ans1}
\eeq
Important point to notice is that $t^3$ term vanishes because it can be written as $\partial_t \la \Omega| \hat{\phi}^2|\Omega\ra$, which must vanish for the Poincare invariant vacuum.

Let us try renormalizing parameters at hand, in such a way that makes \eqref{ans1} finite at 1-loop order. Curiously enough, the finite part of \eqref{ans1} matches the classical evolution exactly at each order in $t$; we have checked this statement explicitly up to $t^7$ corrections. In order to avoid the accidental mismanagement of classical nonlinearities, we perform the loop expansion by keeping $\lambda$ finite but treating the loop factor $\frac{\lambda}{16\pi^2}$ to be small. As a result, in dimensional regularisation, \eqref{ans1} reduces to
\beq
\la C| \hat{\phi}(x) |C\ra (t)\Big |_{\rm 1-loop}=&&\phi_0-\frac{t^2}{2}\left[ \phi_0m_{\rm ph}^2+\frac{\lambda}{3!}\phi_0^3\right]\nonumber\\
&&+\frac{t^4}{4!}\left[\phi_0 m_{\rm ph}^4 +\frac{2\lambda }{3}\phi_0^3m_{\rm ph}^2+\lambda\phi_0^3\Sigma+\frac{\lambda ^2}{2\cdot 3!}\phi_0^5\right]\,.
 \label{ans2}
\eeq
where $m_{\rm ph}^2\equiv m^2+\Sigma$, we have taken into account that $\la \Omega | (\partial_\mu \hat{\phi})^2+m^2\hat{\phi}^2 |\Omega\ra$=0, in dimensional regularization and have set $Z=1$, since it acquires corrections only starting at 2-loop in $\phi^4$ theory\footnote{Notice that $\lambda$ and $Z$ always appear in $\lambda Z$ combination here.}. Also, in this expression $\Sigma$ denotes the usual 1-loop bubble integral
\beq
\Sigma\equiv\frac{\lambda}{2}\int \frac{d^4k}{(2\pi)^4}\frac{-i}{k_\mu^2+m^2}\,.
\label{eq:sigma}
\eeq
In other words, 1-loop computation entails keeping terms linear in $\Sigma$ only.

The key observation we would like to stress on, concerns the initial field acceleration. Being given by the $t^2$-term of \eqref{ans1}, it seems to be determined in terms of the renormalized mass and the bare coupling constant. The latter being infinite in the standard renormalization prescription, our result seems quite strange. One would expect the renormalized coupling to be the one driving the acceleration from the initial value of the field. In order to make the point even clearer, let us simply take Hamilton's operator equation (in the Heisenberg picture) and find its expectation value in the coherent state \eqref{initstate}, at the initial time,
\beq
\Big[\frac{\partial^2}{\partial t^2} \la C| \hat{\phi} |C \ra-\Delta \la C| \hat{\phi} |C \ra +m^2 \la C| \hat{\phi} |C \ra+\frac{\lambda Z}{3!} \la C| \hat{\phi}^3 |C \ra\Big]_{t=0}=0\,.
\eeq
The second term vanishes due to the homogeneity of the coherent state and the vanishing of the one-point vacuum expectation value. The last two terms also simplify to give a result identical to the one deduced from \eqref{ans1}
\beq
\frac{\partial^2}{\partial t^2} \la C| \hat{\phi} |C \ra\Big|_{t=0}=-\phi_0\left(m^2+\frac{\lambda Z}{2}\la\Omega |\hat{\phi}(x)^2 |\Omega\ra\right)-\frac{\lambda Z}{3!}\phi_0^3\,.
\label{inita}
\eeq
It must be emphasized that this expression is exact up to this point. Even if one could question the validity of the time-series expansion adopted earlier, Hamilton's operator equation serves as a core of any perturbative calculation in quantum field theory.

The surprising feature of \eqref{inita} lies in the fact that the initial acceleration of the field value is driven by the bare coupling constant, rather than the one running as a function of $\phi_0$. In other words, the effective potential felt by the field at $t=0$ lacks the logarithmic correction encountered in the celebrated Coleman-Weinberg's work \cite{Coleman:1973jx}. This may not sound too surprising as the latter deals with constant backgrounds, while here we are dealing with an oscillating one (although at the initial moment $\dot{\phi}=0$). In general, the  1-loop effective action can be organised in a derivative expansion
\beq
\Gamma[\phi]=\int d^4 x \Big(-V(\phi^2)-F(\phi^2)(\partial_\mu\phi)^2+P((\partial_\mu\phi)^2)+\ldots\Big)\,,
\label{GammaDer}
\eeq
with ellipsis standing for terms with more than one derivative per field. For constant and homogeneous field configurations one legitimately assumes tree-level values for $F$ and $P$, while including loop effects in the potential. For time-dependent backgrounds, on the other hand, one needs to be more cautious. For instance, it is straightforward to show that up to $\sim\lambda^2$ terms, at 1-loop,  one gets 
\beq
F\sim 1+ \lambda^2\frac{\phi^{2}}{m^{2}}\,\qquad \text{and}\qquad P\sim\frac{\lambda^2}{m^4}(\partial_\mu\phi)^4\,;
\eeq
see \cite{Iliopoulos:1974ur,Fraser:1984zb} for the relevant discussion. Schematic equations for the homogeneous field that follow, take the following form
\beq
\left(1+\frac{\lambda^2 \phi^2}{m^2}+\frac{\lambda^2 \dot{\phi}^2}{m^4}  \right)\ddot{\phi}=-V'(\phi^2)\phi-\lambda^2\frac{\phi}{m^2}\dot{\phi}^2\,.
\label{schematiceq}
\eeq
Although the effective potential depends on renormalized running-coupling, bringing the expression in parentheses to the right-hand-side may significantly affect it. In other words, some of the derivative terms of the effective action modify the background field dynamics at the same level as the radiatively generated potential terms\footnote{Let us stress that this statement holds only in cases when tree-level potential gives a leading contribution to the dynamics and one is interested in sub-leading quantum effects. There are supersymmetric inflationary scenarios in which the inflaton field rolls solely due to radiative corrections to the potential \cite{Dvali:1994ms,Dimopoulos:1997fv}. In those cases the derivative corrections to the effective action give a negligible contribution compared to the quantum potential terms. We would like to thank Gia Dvali for the discussion on this matter.}. In a similar fashion, if we were to consider a configuration with initial $\dot{\phi}$, the last term on the right hand side of \eqref{schematiceq} could give vanishing contribution to $\ddot{\phi}$, together with the combination of mass-term and the third term in parentheses.  

It must be mentioned that the full expression for $F$ obtained in \cite{Iliopoulos:1974ur,Fraser:1984zb} does not seem to lead to the reconciliation with our result for coherent states. In fact, being of semi-classical nature, there is no reason to expect the effective field-acceleration from \eqref{GammaDer} to reproduce the fully quantum result. It must simply serve as a lesson, not to draw conclusions for the dynamical problem based on the static-background analysis. More detailed discussion of these points from the effective action perspective is beyond this work and will be presented elsewhere.

Here, we can shed some light on the possibility of the above-mentioned idea of potential cancelations among derivative terms within the coherent state formalism by introducing non-vanishing initial momentum for the field. For this, let us consider the following initial state
\beq
|C\ra=e^{-i\hat{f}}|\Omega\ra\,, \quad \text{with} \quad \hat{f}\equiv \int d^3 x \left( \phi_{0}\hat{\pi}(x,0)-\pi_{0}\hat{\phi}(x,0) \right)\,.
\eeq
Assuming $Z=1$ for notational simplicity and keeping in mind that it could be easily reintroduced, we get the following extension of \eqref{ans1}
\beq
\la C| \hat{\phi}(x) |C\ra (t)=\phi_0+\pi_0 \cdot t-\frac{t^2}{2}\left[ \phi_0\left(m^2+\frac{\lambda }{2}\la\Omega |\hat{\phi}(x)^2 |\Omega\ra\right)+\frac{\lambda }{3!}\phi_0^3\right]\nonumber \\
-\frac{t^3}{3!}\left[ \pi_0\left(m^2+\frac{\lambda }{2}\la\Omega |\hat{\phi}(x)^2 |\Omega\ra\right)+\frac{\lambda }{2}\phi_0^2\pi_0\right]+\mathcal{O}(t^4)\,.
\eeq
The interesting property of this relation is that the initial momentum does not enter at $t^2$ order, which seems to reassure the statements made around \eqref{schematiceq}.

Closing the bracket on this somewhat speculative attempt for the interpretation of our field-acceleration result, let us go back to \eqref{ans1} and think about an alternative way around the puzzle of having the bare coupling constant being in control of the initial evolution. Splitting the infinite bare coupling into the physical one and a counter-term, as per usual $\lambda=\lambda_{\rm ph}+\delta \lambda$, one can easily convince oneself that the only possibility of getting rid of the infinite leftover term $\left[-\frac{t^2}{2}\frac{\delta \lambda}{3!}\phi_0^3\right]$ is in combination with higher order terms in $t$. For example, there is an infinite term of the form $\left[\frac{t^4}{4!}\frac{\delta \lambda^2}{2\cdot 3!}\phi_0^5\right]$ descending from the last $t^4$-term of \eqref{ans1}, along with other similar ones. The question is if they can be combined into a function of the combination $[\delta \lambda \phi_0^2 t^2]$, vanishing in the infinite limit for the latter. We will tackle this question of time-series resummation within the interaction-picture formalism in section \ref{interactionpicture}.

\subsection{Curious Case of Two-Point Function}

Finally we come to the 2-point correlation function calculated in the coherent state
\beq
\la C(t=0)|\hat{\phi}(x,t)\hat{\phi}(y,0)|C(t=0)\ra=\phi_0^2+\la \Omega|\hat{\phi}(x,0)\hat{\phi}(y,0)|\Omega\ra+t \la \Omega|\hat{\pi}(x,0)\hat{\phi}(y,0)|\Omega\ra\nonumber \\
 -\frac{t^2}{2}\left[ \phi_0^2\left( m^2+\frac{\lambda}{2} \la\Omega|\hat{\phi}^2|\Omega\ra+\frac{\lambda}{3!}\phi_0^2\right)+\la \Omega|\hat{\phi}(y,0)\left(-\Delta+m^2+\frac{\lambda}{2}\phi_0^2+\frac{\lambda}{3!}\hat{\phi}^2(x)\right)\hat{\phi}(x,0)|\Omega\ra \right]\nonumber \\
+\mathcal{O}(t^3)\,.~~
\label{2point}
\eeq
Here, the notation $|C(t=0)\ra$ has been invoked to underline the fact that the coherent state has been constructed using operators at $t=0$, as we are using the Heisenberg picture.

Few interesting comments are in order here:
\begin{itemize}

\item The first two terms of \eqref{2point} should be compared to the result of the background field method. In our case, the second term is a vacuum expectation value, while in the background field method the latter gets replaced with a two-point function for perturbations; the dispersion relation of which depends on the background configuration.

\item On the second line of \eqref{2point}, the first term looks encouraging as it is precisely the combination we have dealt with in \eqref{ans1}. In other words, the renormalization prescription that makes $t^2$-term of \eqref{ans1} finite takes care of that combination here as well.

\item The second term of the second line of \eqref{2point} is also quite appealing, as the operator there resembles a background dependent operator appearing in the equation of motion for perturbations. It must be stressed that this result is reminiscent of the general results regarding Heisenberg picture operators; see section \ref{heisenbergpicture}.

\end{itemize}
\vskip 20pt
Furthermore, we can repackage \eqref{2point} in a simpler form by means of Hamilton's equation as
\begin{flalign}
&\la C(t=0)|\hat{\phi}(x,t)\hat{\phi}(y,0)|C(t=0)\ra=\phi_0^2+\la \Omega|\hat{\phi}(x,t)\hat{\phi}(y,0)|\Omega\ra\nonumber \\
 &~~~~~~~~~-\frac{t^2}{2}\left[ \phi_0^2\left( m^2+\frac{\lambda}{2} \la\Omega|\hat{\phi}^2|\Omega\ra+\frac{\lambda}{3!}\phi_0^2\right)+\frac{\lambda}{2}\phi_0^2\la \Omega|\hat{\phi}(x,0)\hat{\phi}(y,0)|\Omega\ra \right]
+\mathcal{O}(t^3)\,.
\label{2point1}
\end{flalign}
The first two terms are self-explanatory, radiative corrections must be taken care of as usual in order to make the second term finite. Namely, at 1-loop the mass renormalization must be the standard one. As a result, examining the second line of \eqref{2point1} and comparing it with the 1-point function case, we realize that the divergent part of the bare coupling, appearing on the second line of \eqref{2point1}, needs to be resummed with higher order $t$-terms in a similar fashion (see the argument at the end of section \ref{fieldexpval}).

\section{Interaction Picture}
\label{interactionpicture}

In previous sections we have tackled the problem of evolution from a relatively unconventional side. That is, we have chosen the time interval as a primary parameter for organising the expansion, instead of the coupling. One of the advantages of that approach is that it is quite straightforward to account for classical nonlinearities through the underlying quantum process.

In this section we would like to pursue a more standard avenue, involving the expansion in coupling from the get-go but keeping the time-interval of evolution arbitrary. As we are about to show, this corresponds to the resummation of the time-expansion of section \ref{fieldexpval} at a given order in the coupling constant.

Focusing on the homogeneous coherent state \eqref{initstate} studied earlier, we are after the evolution of one-point expectation value
\beq
\la C| \hat{\phi}(t,x) |C\ra\,.
\label{eq:tadpole}
\eeq
Using the standard method we can convert this into the following out-of-time-ordered vacuum expectation value in the interaction picture
\beq
\la C| \hat{\phi}(t,x) |C\ra=\lim_{T\rightarrow \infty}\frac{\la 0| U(T,0)e^{i\phi_0\int d^3x' \hat{\pi}_{I}(0,x')}U(0,t) \FI (t,x) U(t,0)e^{-i\phi_0\int d^3x'' \hat{\pi}_{I}(0,x'')}U(0,-T)|0 \ra}{\la 0|U(T,-T) |0\ra}\,.\nonumber
\label{eq:perturbative_tadpole}
\eeq
Here the subscript '$I$' denotes the corresponding operator to be in the interaction picture, possessing the standard free-particle expansion in ladder operators, '$|0\ra$' stands for the free-theory vacuum and $U$ is the usual time-ordered exponential of the interaction Hamiltonian, i.e. $U(t,t_0)\equiv T\{{\rm exp}(\int_{t_0}^{t}dt~H_{I})\}$; with $H_{I}\equiv \lambda/4!\int d^3z\FI ^4(t,z)$. As it stands, the expectation value in a coherent state has been recast as a vacuum expectation value for asymptotic states with some operator insertions; with the information about the state being imprinted in the exponential field-displacement operators. The latter have a simple effect on enclosed operators, that follows from the Baker--Campbell--Hausdorff identity,
\beq
e^{i\phi_0\int d^3x' \hat{\pi}_{I}(0,x')}\FI(t,x)e^{-i\phi_0\int d^3x'' \hat{\pi}_{I}(0,x'')}=\phi_0 {\rm cos}(mt)+\FI(t,x) \,;
\eeq
where the first term represents a homogeneous solution to free equations of motion, with desired initial conditions.

We evaluate the above expression for the one-point function explicitly  up to $\lambda^2$ order in appendix D. Since in this work we are primarily interested in 1-loop results, let us drop higher order contributions and present a clean result valid up to quadratic order in classical nonlinearities and to the first order in the loop factor (i.e. in $\hbar$, if we were to reinstate it)
\beq
\la C| \hat{\phi}(t,x) |C\ra_{1-loop}=&&\phi_0{\rm cos}(m_{\rm ph}t)-\frac{\lambda \phi_0^3}{3!}\frac{{\rm sin}(m_{\rm ph}t)(6m_{\rm ph}t+{\rm sin}(2m_{\rm ph}t))}{16m_{\rm ph}^2}\nonumber\\
&&+\frac{\lambda^2\phi_0^5}{4!^22^6m^4}\Big( (23-72m^2t^2){\rm cos}(mt)-24~{\rm cos}(3mt)+{\rm cos}(5mt)\nonumber \\
&&~~~~~~~~~~~+12~mt(8~{\rm sin}(mt)-3~{\rm sin}(3 mt)) \Big)\nonumber\\
&&+\frac{\lambda^2}{2}\int_0^t dt_1\int_{0}^{t_1}dt_2\int\frac{d^3p}{(2\pi)^3}\frac{{\rm sin}(2E_p(t_1-t_2))}{(2E_p)^2}\frac{{\rm sin}(m(t-t_1))}{m}\Phi(t_1)\Phi(t_2)^2\nonumber\\
&&+\mathcal{O}(\lambda^3)\,,
\label{finint}
\eeq
where, $\Phi(t)\equiv \phi_0 {\rm cos}(mt)$ and $m_{\rm ph}$ is defined in the usual manner as
\beq
m_{\rm ph}^2\equiv m^2+\frac{\lambda}{2}\la \FI^2 \ra\,.
\eeq
Here the second term is the expectation value in a free theory vacuum, a.k.a. 'Bubble Diagram'. Notice that one can use $m$ and $m_{\rm ph}$ interchangeably in \eqref{finint} everywhere except for the first line, due to the order of approximation.

Interestingly, and expectantly, our calculation recovers the classical nonlinear evolution through quantum dynamics, supplemented with quantum corrections. Namely, the first three terms are precisely what one would obtain by solving the classical homogeneous equation perturbatively in $\lambda$. The last term, on the other hand, is a 1-loop divergent contribution. It is therefore natural to expect this divergence to renormalize the coupling constant, as the mass renormalization has been already taken care of. Neglecting this prejudice, the fate of the divergence should be determined by its time dependence, which we are about to check.

Naively, the momentum integral in the last term of \eqref{finint} appears to be linearly divergent, which is not the behaviour we would expect for the one-loop diagram responsible for the coupling constant renormalization. However, such assessment fails due to the appearance of the oscillatory function of energy. After swapping momentum and time integrals and explicitly performing the latter, it becomes evident that the divergence is logarithmic.

In order to verify that the computation of the current section is up to this point compatible with the time-expansion analysis of previous sections, we may blatantly expand \eqref{finint} in Taylor series to $t^4$ order (preceded by explicit integration over $t_{1,2}$). It is a matter of straightforward computation to show that this leads to the result identical to \eqref{ans2}. In other words, the last term of \eqref{finint} is $\mathcal{O}(t^4)$, in time-expansion, and thus does not contribute to the initial field-acceleration. 
For the latter we once again arrive at
\beq
\lim_{t\rightarrow 0}\partial_t^2 \la C| \hat{\phi}(t,x) |C\ra_{1-loop}=-m_{\rm ph}^2\phi_0-\frac{\lambda}{3!}\phi_0^3\,.
\label{eq:eff_pot}
\eeq
For the physical discussion and potential interpretation of this result, the reader is redirected to section \ref{fieldexpval}.

At this point one may still wonder, whether in the current interaction-picture computation, the time-expansion of the integrand in the last term of \eqref{finint} is legitimate (due to the fact that every single term in this expansion seems to diverge). Instead, let us proceed by isolating the divergent part and then compute the leftover finite contribution without time-expansion. For this we use the trick that has been used before in a similar context in \cite{calzetta}. In particular, in order to rewrite the last term of \eqref{finint} in a manifestly logarithmically divergent form, we rewrite the momentum-dependent oscillatory term as
\beq
\sin{\left(2E_p(t_1-t_2)\right)} = \frac{1}{2E_p}\frac{d}{dt_2}\cos{\left(2 E_p(t_1-t_2)\right)}\,.
\label{bp}
\eeq
Integration of this by parts in the integrand of \eqref{finint} leads to the emergence of three terms: two boundary terms and a term with derivative acting on $\Phi(t_2)$. The upper boundary term, after further integration over $t_1$ and remembering that at this order in our expansion we can replace $m$ with $m_{\rm ph}$, gives the following contribution 
\beq
\la C|\hat\phi(t,x)|C\ra\supset \frac{\lambda ^2 \phi _0^3}{2}\frac{\sin \left( m_{\text{ph}}t\right) \left(6 
   m_{\text{ph}}t+\sin \left(2  m_{\text{ph}}t\right)\right)}{16\, m_{\text{ph}}^2}\int \frac{d^3p}{(2\pi)^3}\,\frac{1}{(2E_p)^3}\,.
\eeq
Interestingly, this term exhibits the time-dependence identical to the second term of \eqref{finint}, with the momentum integral corresponding to the 1-loop diagram responsible for the coupling constant renormalization in scattering computations. Therefore, the contribution in question can be absorbed by the term linear in $\lambda$ as a coupling renormalization
\beq
\lambda_{\rm ph}=\lambda-3\lambda^2 \int \frac{d^3p}{(2\pi)^3}\,\frac{1}{(2E_p)^3}\,.
\eeq

\begin{figure}

\hskip 50pt \includegraphics[width = 0.75\linewidth]{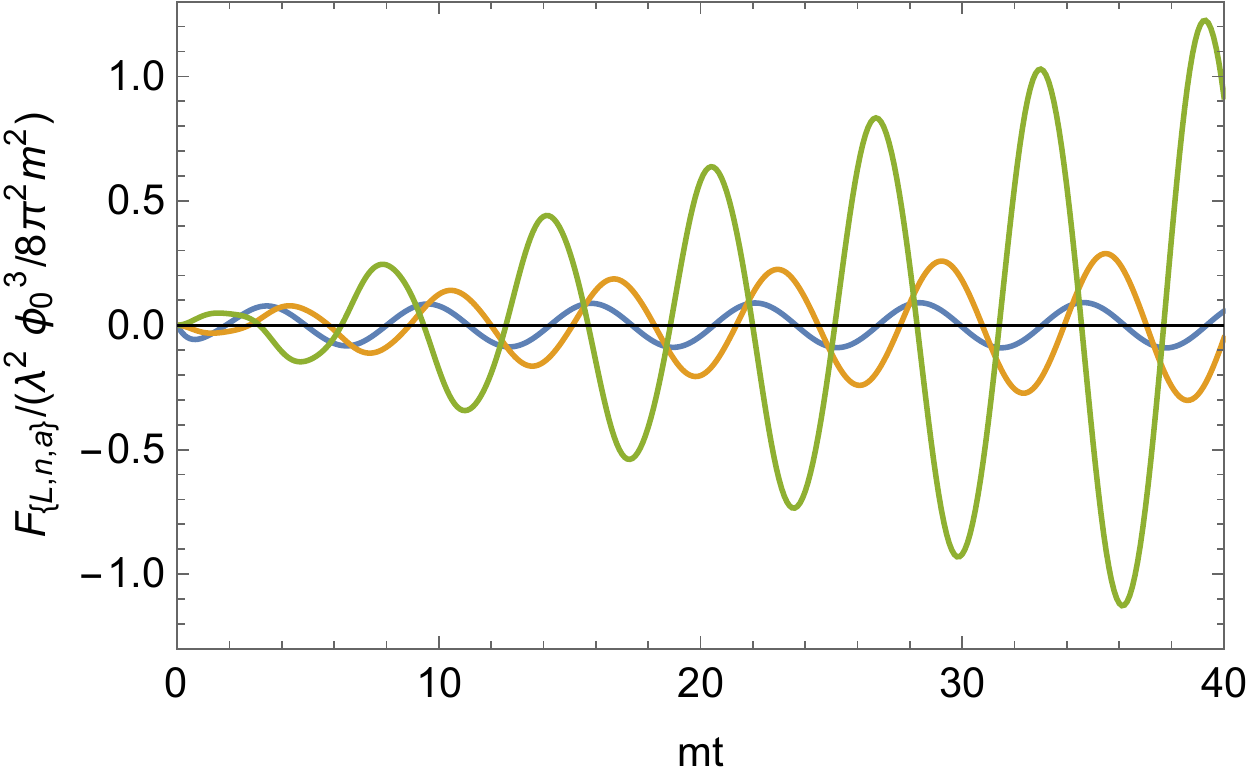}
\caption{\small Numerical comparison of finite quantum corrections descendent from the last term of \eqref{finint}. The \textit{blue} curve denotes $F_L$, the \textit{orange} one corresponds to $F_n$ and the $green$ one represents $F_a$.}

\label{meijer}

\end{figure}

The remarkable outcome of the adopted trick lies in the fact that the other two terms resulting from integration-by-parts are finite. The lower boundary term (henceforth denoted as $F_{\rm L}$) descending from the last line of \eqref{finint}, gets expressed through Meijer G-functions. The numerical behaviour of this contribution has been depicted on Fig. \ref{meijer}. Away from the initial moment, it seems to undergo steady oscillations. However, it exhibits peculiar behaviour near $t=0$, namely
\beq
\lim_{t\rightarrow 0}F_{\rm L}(t)=\frac{\lambda^2\phi_0^3}{64 \pi^2}\left( -\frac{1}{2}+\gamma+{\rm ln }\left( mt\right) \right)t^2+\mathcal{O}(t^4)\,;
\label{initsing}
\eeq
which implies that the contribution to the field-acceleration, i.e. $\partial_t^2 \la C| \hat{\phi}(t,x) |C\ra$, has a logarithmic initial time singularity. This is believed to be an artefact of the weak coupling expansion; see \cite{calzetta} and references therein. Moreover, the unusual behaviour we have encountered in the time-expansion analysis is related to $F_{\rm L}(t)$-contribution as well. Namely, if we were to expand the integrand of $F_{\rm L}(t)$ in a Taylor series in $t$, before momentum integration, we would get a logarithmically UV-diverging  $t^2$-term (which would undo the coupling renormalization) and quadratically UV-diverging $t^4$-term. In other words, the peculiar initial time behaviour exhibited by \eqref{initsing} seems to be the price for recovering the standard renormalization prescription.

The remaining term resulting from integrating \eqref{bp} in the last term of \eqref{finint}, consist of two parts. The part that can be analytically calculated (which we denote as $F_a(t)$) and the one requiring numerical integration (denoted by $F_n(t)$). Together with $F_{\rm L}$, they are depicted in Fig. \ref{meijer}, demonstrating that eventually $F_a$ dominates over the other contributions. The analytic expression for $F_a(t)$ takes the following form
\beq
F_a(t)=\frac{\lambda^2\phi_0^3}{32\pi^2}\frac{{\rm sin}(mt)(2mt+{\rm sin}(2mt))}{16m^2}\,.
\eeq

Putting everything together, \eqref{finint} can be written as
\beq
\la C| \hat{\phi}(t,x) |C\ra_{1-loop}=&&\phi_0{\rm cos}(m_{\rm ph}t)-\frac{\lambda_{\rm ph} \phi_0^3}{3!}\frac{{\rm sin}(m_{\rm ph}t)(6m_{\rm ph}t+{\rm sin}(2m_{\rm ph}t))}{16m_{\rm ph}^2}\nonumber\\
&&+\frac{\lambda_{\rm ph}^2\phi_0^5}{4!^22^6m_{\rm ph}^4}\Big( (23-72m_{\rm ph}^2t^2){\rm cos}(m_{\rm ph}t)-24~{\rm cos}(3m_{\rm ph}t)+{\rm cos}(5m_{\rm ph}t)\nonumber \\
&&~~~~~~~~~~~+12~m_{\rm ph}t(8~{\rm sin}(m_{\rm ph}t)-3~{\rm sin}(3 m_{\rm ph}t)) \Big)\nonumber\\
&&+\frac{\lambda_{\rm ph}^2\phi_0^3}{32\pi^2}\frac{{\rm sin}(m_{\rm ph}t)(2m_{\rm ph}t+{\rm sin}(2m_{\rm ph}t))}{16m_{\rm ph}^2}+F_n(t)+F_{\rm L}(t)\nonumber\\
&&+\mathcal{O}(\lambda^3)\,,
\label{finexp}
\eeq

\begin{figure}

\hskip 50pt \includegraphics[width = 0.75\linewidth]{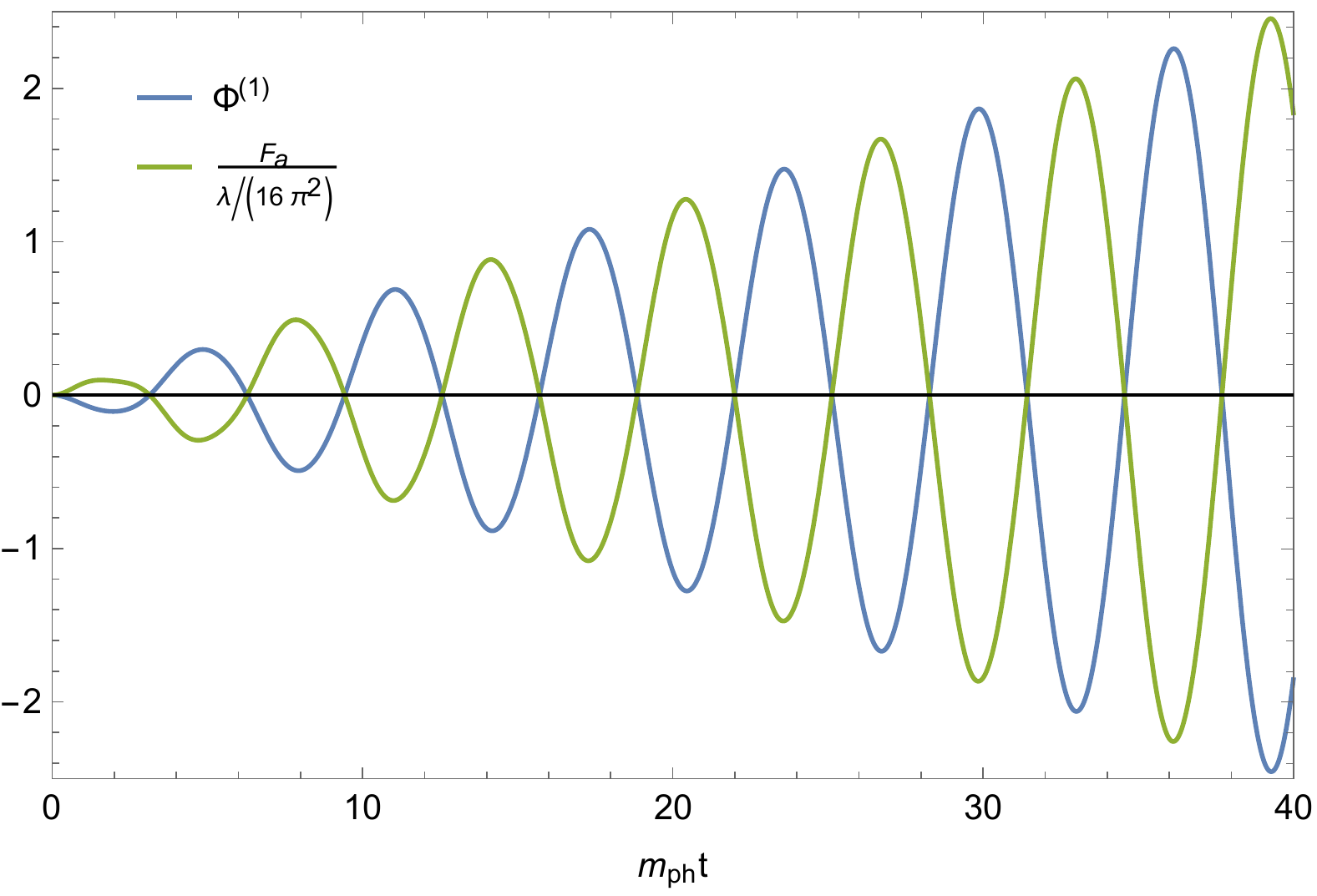}
\caption{\small The \textit{blue} curve denotes the second term of \eqref{finexp}, while the \textit{green} one depicts $F_a$ (rescaled by the loop factor for the visibility). The plot is displayed in units of $\lambda_{\rm ph} \phi_0^3/m_{\rm ph}^2$.}
\label{leadcorr}

\end{figure}

As one can easily notice, besides the renormalization of parameters, there are actual quantum corrections to the classical dynamics. Moreover, the dominant quantum effect can be evaluated analytically and it exhibits significant similarity with the linear classical correction to the dynamics; (ignoring numerical coefficients of order unity) the former is suppressed by the loop factor. Moreover, it is easy to see that these contributions are always out of phase, as depicted in Fig. \ref{leadcorr} for convenience. Due to the linear growth of the correction in question, one might expect it to eventually dominate over the first term of \eqref{finexp}. However, taking into account the secular instability present in classical corrections, which we know for a fact to be an artefact of the approximation, one may wonder whether the observed quantum instability is equally spurious. In fact, at the level of the carried-out computation, the classical evolution breaks down (departs from full nonlinear evolution) around the timescale
\beq
t_{\rm cl}=\left[\frac{\lambda_{\rm ph} \phi_0^2}{16m_{\rm ph}} \right]^{-1}\,.
\eeq
In a similar fashion, if we were to estimate the time it would take for the quantum contributions to become of order $\phi_0$, we would obtain
\beq
t_{\rm q}=t_{\rm cl}\left(\frac{\lambda_{\rm ph}}{16\pi^2}\right)^{-1}\,.
\label{tq}
\eeq
It must be stressed that this timescale matches the expression for the quantum-break-time given in \cite{Dvali:2017eba}, according to which, assuming re-scattering of the background constituents to be the underlying process responsible for quantum breaking, and assuming $2\rightarrow 2$ scattering to be the simplest contributing process (treating particles to be collectively off-shell, thus ignoring the kinematic prohibition of the process) one would get the rate of quantum depletion to be given by $N/t_{\rm q}$ where $N$ is the number of particles per Compton-volume.

Regretfully, more definite statements require going to higher orders in $\lambda$. In particular, we know for a fact that there are processes that change the total number of particles, the most elementary one being $4\rightarrow 2$, with rate proportional to $\lambda^4$; see the discussion in \cite{Dvali:2017eba}. In other words, we expect the quantum part of \eqref{finexp} to have higher order corrections in the classical coupling $\lambda$. Therefore, it remains an open question whether the observed quantum secular instability survives in the resummed theory.

\section{Summary and Outlook}

In this work, we have studied the real-time quantum dynamics of coherent states in the case of an interacting quantum field theory. One of the vital details about the setup is the build-up over the vacuum of the interacting theory, rather than of the free Hamiltonian. In this way we manage to bypass some of the unnecessary approximations, which in turn helps us to avoid some of the complications that could arise in case of inconsistent perturbative treatment of the dynamics and the state itself.

In order to establish the gradual departure of physical observables from their classical trajectories, we have begun the analysis by organizing the dynamics as a Taylor series in time elapsed since the initial moment. This method seemed particularly promising for computing the initial values of certain physical quantities.  Below we recap some of our central results in this regard.

Considering a coherent state \eqref{cohst} corresponding to the configuration with homogeneous field $\phi_0$ and its momentum $\pi_0$, the initial expectation values of the Hamiltonian density and the field-acceleration take the following form at one-loop\footnote{Notice that the exact results given by \eqref{initiclassham} and \eqref{inita} are equally elegant. However, we refrain from making statements beyond 1-loop here, since we have mainly focused our discussion on the latter.}
\beq
\label{hamden}
&\la C| \hat{\mathcal{H}}(x) |C \ra(t=0)&=\frac{1}{2}\pi_{0}^2+\frac{1}{2}m_{\rm ph}\phi_{0}^2+\frac{\lambda}{4!}\phi_{0}^4\,,\\
&\partial_t^2 \la C| \hat{\phi}(x) |C\ra (t=0)&=-m_{\rm ph}^2\phi_0-\frac{\lambda}{3!}\phi_0^3\,,
\label{facc}
\eeq
where $m_{\rm ph}$ stands for the renormalized mass. The equation \eqref{facc} simply follows from the expectation value of Hamilton's operator equation, evaluated at the initial moment. The unexpected property of these results lies in the fact that these quantities seem to be governed by the tree-level potential (i.e. no evidence for coupling renormalization at this level). In fact, as we have argued qualitatively in section \ref{fieldexpval}, the standard expectation regarding the field-acceleration needs to be reconsidered by including radiatively induced derivative interactions in the effective action, as they seem to contribute to the dynamics as significantly as the potential terms (irrespective of the rate of the dynamics). Unfortunately, after all is said and done the manifestly infinite contributions remain (through the bare coupling) and need to be disposed off via resummation.

Another interesting observation we would like to bring attention to, concerns the initial-time two-point function. In particular, the following result holds to all orders in perturbation theory\footnote{The only property of the theory used in deriving this expression was that the vacuum expectation value of odd number of fields vanishes.}
\beq
\la C| \hat{\phi}(0,x) \hat{\phi}(0,y) |C\ra =\phi_0^2+\la \Omega| \hat{\phi}(0,x) \hat{\phi}(0,y) |\Omega\ra\,.
\eeq
The striking similarity with the background field technique is obvious here. Note however that the second term is a background independent vacuum two-point function, while in the semiclassical treatment it comes out to be the correlator for fluctuations with background-dependent dispersion relation.

Moving forward, we have computed first several terms in the time-series of $\la C| \hat{\phi}(t,x) |C\ra (t)$, illustrated in \eqref{ans1}. It is straightforward to see that the effect of classical nonlinearities is correctly captured by every computed term. In fact, all finite terms obtained in this expansion seem to correspond to classical contributions; this statement has been verified explicitly up to $t^7$ order. We would like to emphasize that the reproduction of classical contributions at each order in $t$ is performed in full generality without resorting to the weak coupling expansion. In this approach, quantum effects seem to appear in the form of vacuum expectation values of singular operators, some of which can be absorbed by the parameters of the theory, while the others need to be resummed with higher order terms in $t$.

Due to the above-mentioned need for the resummation of the time-series, we have performed the calculation of $\la C| \hat{\phi}(t,x) |C\ra (t)$ within the interaction-picture framework. Even though the starting point of every single calculation invokes the infinitesimal time limit of the Hamiltonian flow, the method in question manages the resummation of the time-series at a given order in perturbation theory (in coupling constant), yielding a result valid for extended periods of time. In particular, we have evaluated the quantity in question up to $\lambda^2$ order, further simplifying the result by dropping 2-loop contributions, leading to \eqref{finexp}, which includes nonlinear classical effects up to order $\lambda^2$ as well as 1-loop quantum corrections; with parameters being renormalized as per the standard prescription. The appearance of the standard coupling renormalization, unlike \eqref{facc}, shows potential in terms of recovering the effects of Coleman-Weinberg effective potential to some extent, but for that one needs to extend our 1-loop calculation to higher orders in $\lambda$ (classical coupling) and perform the resummation. The outcome of the latter is especially interesting, due to the appearance of spurious secular divergencies in \eqref{finexp} that are obviously an artefact of the perturbative calculation. The obtained quantum corrections to the dynamics have interesting properties, with the most notable ones being:

\begin{itemize}

\item \textit{Logarithmic initial-time singularity} $$\lim_{t\rightarrow 0}\partial_t^2 \la C| \hat{\phi}(t,x) |C\ra\supset \frac{\lambda^2\phi_0^3}{32\pi^2}{\rm ln}(mt).$$ Similar pathologies have been previously encountered in the literature. Namely, in the background field methods one sometimes considers the initial state of perturbations to be given by the free-theory vacuum for fluctuations leading to initial-time singularities, which is believed to be an artefact of the adopted, strictly speaking ill-defined, treatment of the initial state (see the discussion in \cite{calzetta} and references therein). In our coherent state calculation, on the other hand, we have not invoked any free-theory states or operators in the construction. 
However, we have shown that the perturbative expansion of the interaction picture and subsequent termination of the series at 1-loop effectively reduced our expression to the one we would obtain if the initial coherent state were built from a free theory vacuum.

\item \textit{Growing quantum correction to the classical dynamics.} It has a form similar to the classical secular (spurious) instability, albeit with a loop-factor suppression. Therefore, it is incapable of giving a significant correction to the dynamics; the perturbative expansion breaks down before quantum terms have time to grow sufficiently. However, if taken for the face value and compared to the initial field displacement $\phi_0$, it leads to the estimate of \cite{Dvali:2017eba} for the quantum break-time (i.e. the time it takes for the quantum evolution to depart significantly from the classical dynamics), which was obtained using the simplest $2\rightarrow 2$  re-scattering of the constituents as a source of quantum breaking.

\end{itemize}
\vskip 10pt
We have also studied the quantum evolution of a coherent state for infinitesimal time $t$, and its subsequent overlap with a coherent state constructed out of the classically evolved field configuration. We have computed the probability for a quantum state at time $t$ to coincide with the aforementioned classical counterpart. The deviation is schematically given as 
\beq
|\la C_{cl}|C(t)\ra |^2=1-t^2 D+\mathcal{O}(t^3)\,.
\eeq
The notable fact about this expression is that the probability departs from unity quadratically in $t$, which is due to the fact that the infinitesimal transition amplitude between coherent states is a pure phase at leading order in $t$ (see the discussion of sections \ref{departurerate} and \ref{pathintegral}). Regretfully, the coefficient $D$ is divergent in (3+1)-dimensions and a resummation of higher order terms in $t$ seems to be required. However, in (1+1)-dimensions one gets an elegant result given by \eqref{p2}, which shows that the characteristic timescale of departure from the classical path behaves as $\lambda^{-1}$. This, on the other hand, corresponds to a parametrically shorter time than \eqref{tq}.

It is quite possible for the computed departure probability to capture more of the microscopic dynamics than the one-point function. However, the above-mentioned peculiar result may be an artefact of the infinitesimal-time approximation. In particular, examining the quantum contribution to \eqref{finexp} it is easy to see that, while over timescales longer than the inverse-mass the correction is linear in time (giving $t_{\rm q}\sim \lambda^{-2}$), for infinitesimal time-interval the corrections become quadratic in time thus giving $t_{\rm q}\sim \lambda^{-1}$. In other words, departure rates observed for infinitesimal time may be part of oscillatory behaviour, similar to \eqref{finexp}.

We would like to conclude by briefly mentioning future directions. Our calculation can be straightforwardly extended to include higher order corrections in coupling, which may clarify whether the growing quantum effects (depicted in \eqref{finexp}) are spurious or not. Moreover, as we have already pointed out, going to higher orders in coupling may open up another channel of depletion connected to the particle number changing processes. Alternatively one could introduce another light particle coupled to $\phi$, unlocking the two-particle annihilation channel. It would be interesting to see how the particle production in the adopted coherent state formalism compares to the standard semi-classical analysis which studies the production of particles in the time-dependent classical background. These and other related questions will be discussed elsewhere.

\section*{Acknowledgements}

We would like to thank Gia Dvali for many illuminating discussions and comments on the manuscript. We also thank Giordano Cintia, Cesar Gomez, Georgios Karananas, Otari Sakhelashvili, Dimitris Skliros, Tanmay Vachaspati and Sebastian Zell for valuable discussions. 

\newpage

\section*{Appendix A: Coherent State: Normal Ordering}
\renewcommand{\theequation}{A-\Roman{equation}}
\setcounter{equation}{0}

In this appendix, we would like to discuss a point that could, in principle, be used to question the legitimacy of the adopted construction.
We have considered the following coherent state
\beq
|C\ra=e^{-i\hat{f}}|\Omega\ra\,,\qquad \text{with}\qquad \hat{f}\equiv \int d^3x\phi_0\hat{\pi}\,;
\label{initstate1}
\eeq
focusing on a homogeneous one for simplicity.
This may appear to be problematic, because of the same point singularity. Even though, the state defined this way is properly normalised, i.e. $\la C|C \ra=1$, one should keep in mind that the exponential operator is defined as its Taylor series
\beq
|C\ra=\sum_{n=0}^{\infty}\frac{(-i\phi_0)^n}{n!}\int d^{3}x_1\ldots d^{3}x_n \hat{\pi}(x_1)\ldots \hat{\pi}(x_n)|\Omega\ra\,.
\label{cohexp}
\eeq
Almost every term in this expansion possesses a singularity, for instance the operator
\beq
\int d^{3}x_1d^{3}x_2 \hat{\pi}(x_1)\hat{\pi}(x_2)
\eeq
is singular. Easiest way to convince oneself in this statement is to calculate the vacuum expectation value
\beq
\int d^{3}x_1d^{3}x_2 \la \Omega|\hat{\pi}(x_1)\hat{\pi}(x_2)|\Omega\ra\,,
\eeq
which diverges. It is straightforward to show that to the leading order in perturbation theory it possesses an infrared divergence (i.e. is proportional to the volume of the system). The ultra-violate singularity from the overlapping points, on the other hand, can be regulated. At this point we may wonder whether singularities must be brushed off term-by-term from \eqref{cohexp}, using normal ordering. This is a legitimate point, since there would be no objection against the state constructed using non-singular operators acting on the vacuum. Indeed, let us see what happens when we go down this road. For illustrative purposes, we will perform analysis up to certain order in $\phi_0$, in particular we begin by improving our coherent state as
\beq
|C\ra=\left( 1-i\phi_0\int d^3x_1\hat{\pi}(x_1)+\frac{(-i\phi_0)^2}{2}\int d^3x_1d^3x_2:\hat{\pi}(x_1)\hat{\pi}(x_2):\right.\nonumber\\ \left.+\frac{(-i\phi_0)^3}{3!}\int d^3x_1d^3x_2d^3x_3:\hat{\pi}(x_1)\hat{\pi}(x_2)\hat{\pi}(x_3):\right.\nonumber\\
 \left.+\frac{(-i\phi_0)^4}{4!}\int d^3x_1d^3x_2d^3x_3d^3x_4:\hat{\pi}(x_1)\hat{\pi}(x_2)\hat{\pi}(x_3)\hat{\pi}(x_4):+\ldots\right)|\Omega\ra\,.
\label{cohexp1}
\eeq
As one can see, we have simply replaced operators with their normal ordered versions. See \cite{Skliros:2015vpa,Ellis:2015xwp} for the discussion on complete normal ordering. For convenience, we give the explicit expressions that will be of use
\beq
&&:\hat{\pi}(x_1)\hat{\pi}(x_2):=\hat{\pi}(x_1)\hat{\pi}(x_2)-\la\Omega| \hat{\pi}(x_1)\hat{\pi}(x_2)|\Omega\ra\,,\\
&&:\hat{\pi}(x_1)\hat{\pi}(x_2)\hat{\pi}(x_3):=\hat{\pi}(x_1)\hat{\pi}(x_2)\hat{\pi}(x_3)-3\hat{\pi}(x_1)\la\Omega|\hat{\pi}(x_2)\hat{\pi}(x_3)|\Omega\ra\,,\\
&&:\hat{\pi}(x_1)\hat{\pi}(x_2)\hat{\pi}(x_3)\hat{\pi}(x_4):=\hat{\pi}(x_1)\hat{\pi}(x_2)\hat{\pi}(x_3)\hat{\pi}(x_4)-6:\hat{\pi}(x_1)\hat{\pi}(x_2):\la\Omega|\hat{\pi}(x_3)\hat{\pi}(x_4)|\Omega\ra\nonumber \\
&&\hskip 135pt-\la\Omega|\hat{\pi}(x_1)\hat{\pi}(x_2)\hat{\pi}(x_3)\hat{\pi}(x_4)|\Omega\ra\,.
\eeq
Here, we have implicitly assumed that both sides are integrated over coordinates. Although we have gotten rid of singularities, we need to check the normalization. It is easy to see that \eqref{cohexp1} does not have a unit norm. Upon normalization we get
\beq
|C\ra=&&\Big( 1-\frac{\phi_0^2}{2}\int d^3y_1d^3y_2\la \Omega |\hat{\pi}(y_1)\hat{\pi}(y_2)|\Omega\ra\nonumber \\
&&+\frac{\phi_0^4}{24}\int d^3y_1d^3y_2d^3y_3d^3y_4\la \Omega |\hat{\pi}(y_1)\hat{\pi}(y_2)\hat{\pi}(y_3)\hat{\pi}(y_4)|\Omega\ra +\ldots\Big)\nonumber \\
&&\times\left( 1-i\phi_0\int d^3x_1\hat{\pi}(x_1)+\frac{(-i\phi_0)^2}{2}\int d^3x_1d^3x_2:\hat{\pi}(x_1)\hat{\pi}(x_2):\right.\nonumber\\ 
&&\left.+\frac{(-i\phi_0)^3}{3!}\int d^3x_1d^3x_2d^3x_3:\hat{\pi}(x_1)\hat{\pi}(x_2)\hat{\pi}(x_3):\right.\nonumber\\
 &&\left.+\frac{(-i\phi_0)^4}{4!}\int d^3x_1d^3x_2d^3x_3d^3x_4:\hat{\pi}(x_1)\hat{\pi}(x_2)\hat{\pi}(x_3)\hat{\pi}(x_4):+\ldots\right)|\Omega\ra\,.
\label{cohexp2}
\eeq
The first line represents the normalization factor. It is straightforward to show that upon expansion this normalization factor simply removes the normal ordering and reduces \eqref{cohexp2} to \eqref{cohexp}. In conclusion we are back to square one.

For non-interacting case, one can easily re-sum the normalization factor. In fact, the normalized normal ordered state in that case reduces to
\beq
|C\ra=e^{-\frac{\phi_0^2}{2}\int d^3x_1 d^3x_2\la \hat{\pi}(x_1)\hat{\pi}(x_2) \ra}:e^{-i\phi_0\int d^3x\hat{\pi}(x)}:|\Omega\ra\,.
\eeq
This is precisely what one gets in quantum mechanics. Expanding in terms creation-annihilation operators we would arrive at the familiar expression.

In conclusion, the imposition of normal ordering brings the norm away from $1$, which can be fixed by introducing a constant normalization factor, amounting to the removal of the normal ordering.

\section*{Appendix B: Path-Integral: Green's Functions}
\renewcommand{\theequation}{B-\Roman{equation}}
\setcounter{equation}{0}

Here, we would like to discuss a generating functional for Green's functions within the path-integral formalism overviewed in section \ref{pathintegral}. In other words, we are interested in a path-integral representation of the transition amplitude with field insertions
\beq
\la\phi_{out},\pi_{out};t_{out}|\prod_k\hat{\phi}(x_k,t_k)|\phi_{in},\pi_{in};t_{in}\ra\,, \quad \text{with} \quad t_k>t_{k+1},~~\forall ~k\,.
\eeq
As usual, we will dissect the time interval into infinitesimal ones by inserting the resolution of identity, as a result some of the segments will add up with a field operator sandwiched inside. Having already discussed segments without insertions in section \ref{pathintegral}, let us focus on the one with a field insertion
\beq
\la\phi'_c,\pi'_c;t+dt|\hat{\phi}(x,\tau)|\phi_c,\pi_c;t\ra \quad \text{with}\quad \tau\in [t,t+dt]\,.
\eeq
This element can be straightforwardly evaluated to the linear order in $dt$ and field variations as
\beq
\lim_{dt\rightarrow 0}\la\phi'_c,\pi'_c;t+dt|\hat{\phi}(x,\tau)|\phi_c,\pi_c;t\ra=\left[ 1-idt\bar{H}[\phi_c,\pi_c] \right]\phi_c(x,t)-idt\int d^3z \Big(\dot{\pi}_c(z)\la\hat{\phi}(z)\hat{\phi}(x)\ra\nonumber \\
+\partial_{\vec{z}}\phi_c(z)\la\partial_{\vec{z}}\hat{\phi}(z)\hat{\phi}(x)\ra+m^2\phi_c(z)\la \hat{\phi}(z)\hat{\phi}(x) \ra+\frac{\lambda}{3!}\phi_c^3(z)\la \hat{\phi}(z)\hat{\phi}(x)\ra+\frac{\lambda}{3!}\phi_c(z)\la \hat{\phi}^3(z)\hat{\phi}(x) \ra\Big)\,;
\eeq
where, $\la\ldots\ra$ stands for the vacuum expectation value. The second line can be simplified further using Heisenberg's operator equation equation of motion, but it's unnecessary. Notice that we will, once again, pile-up intervals of this kind together. As a result, the infinitesimal term $dt\,\bar{H}$ will turn into an exponentiated time-integral. The infinitesimal term involving expectation values, on the other hand, receives only contributions from intervals with field insertions and thus vanishes in the limit $dt\rightarrow 0$.

Therefore, it seems one can make the case for the familiar path-integral representation for Green's functions (as stated in \cite{Zhang:1999is})
\beq
\la\phi_{out},\pi_{out};t_{out}|\prod_k\hat{\phi}(x_k,t_k)|\phi_{in},\pi_{in};t_{in}\ra \simeq \int [d\mu(\phi,\pi)]\prod_k\phi(x_k,t_k)e^{i\int_{t_{in}}^{t_{out}} dt\left(\int d^3x \dot{\phi}\pi-\bar{H}\right)}\nonumber \\
\times e^{i\int d^3x \left[\pi_{out}\phi_{out}-\pi_{in}\phi_{in}\right]}\,.
\label{greenpath}
\eeq
From this, one can define the generating functional and quantum effective action as usual, through the Legendre transform.

Next, we could derive Coleman-Weinberg potential using the usual background field method; underlining the fact that all the complications we have encountered with infinites in \eqref{dd} are hidden in '$\simeq$' sign appearing in \eqref{greenpath}.

\section*{Appendix C: Expectation Value of the Field Operator}
\renewcommand{\theequation}{C-\Roman{equation}}
\setcounter{equation}{0}

Here we give expressions for the relevant commutation relation used in section \ref{fieldexpval} to evolve the expectation value of the field operator using the relations given in sections \ref{heisenbergpicture} and the Baker--Campbell--Hausdorff formula. In particular
\beq
&&[\hat{H},\hat{\phi}(x)]=-\frac{i}{Z}\hat{\pi}\,,\\
&&[\hat{H},[\hat{H},\hat{\phi}(x)]]=-\Delta \hat{\phi}(x)+m^2\hat{\phi}+\frac{\lambda Z}{3!}\hat{\phi}^3\nonumber\\
&&~~~~~~~~~~~~~~~~~~~~~~~~~~~~~~+\int d^3z \vec{\nabla}^{(z)}\left( \delta^{(3)}(x-z)\vec{\nabla}^{(z)}\hat{\phi}(z)\right)\,,
\eeq
(The last term is a boundary term, the likes of which we will be dropping; by assuming that the observable will be evaluated away from the boundary.)
\beq
&&[\hat{H},[\hat{H},[\hat{H},\hat{\phi}(x)]]]= -\frac{i}{Z}\left[ (-\Delta+m^2)\hat{\pi}+\frac{\lambda Z}{4}\left( \hat{\pi}\hat{\phi}^2+\hat{\phi}^2\hat{\pi} \right) \right]\,,\\
&&[\hat{H},[\hat{H},[\hat{H},[\hat{H},\hat{\phi}(x)]]]]= (-\Delta+m^2)^2\hat{\phi}+\frac{\lambda Z}{3!}(-\Delta+m^2)\hat{\phi}^3+\frac{\lambda Z}{2}\hat{\phi}^2(-\Delta+m^2)\hat{\phi}\nonumber \\
&&~~~~~~~~~~~~~~~~~~~~~~~~~~-\frac{\lambda}{4Z}\left( \hat{\pi}^2\hat{\phi}+2\hat{\pi}\hat{\phi}\hat{\pi}+\hat{\phi}\hat{\pi}^2 \right)+\frac{(\lambda Z)^2}{2\cdot 3!}\hat{\phi}^5\,.
\eeq

We have used these expressions, together with the fact that the commutators of shifted operators (appearing in \eqref{heisen}) obey similar relations, in deriving \eqref{ans1}.

\section*{Appendix D: Interaction-Picture Computation}
\renewcommand{\theequation}{D-\Roman{equation}}
\setcounter{equation}{0}

Here, we would like to give some of the intermediate steps used in arriving at the results of section \ref{interactionpicture}. In a nutshell, this appendix represents the recount of the standard interaction-picture steps applied to the evaluation of \eqref{eq:perturbative_tadpole}.

The operators $\FI$ and $\hat{\pi}_{I}$, appearing in section \ref{interactionpicture}, are the usual interaction picture operators
\beq
&&\FI (t,x)=\int \frac{d^3 p}{(2\pi)^3}\frac{1}{\sqrt{2 E_p}}\left( \hat{a}_pe^{ip_\mu x^\mu}+\hat{a}^{\dagger}_pe^{-ip_\mu x^\mu} \right)\Big|_{x^0=t-t_0}\,,\\
&&\hat{\pi}_I (t,x)=\int \frac{d^3 p}{(2\pi)^3}(-i)\sqrt{\frac{E_p}{2}}\left( \hat{a}_pe^{ip_\mu x^\mu}-\hat{a}^{\dagger}_pe^{-ip_\mu x^\mu} \right)\Big|_{x^0=t-t_0}\,,
\eeq
with $p^0=E_p\equiv\sqrt{m^2+\vec{p}^2}$ and $t_0$ being a fiducial moment of time at which the ladder operators have been defined (not to be confused with moment of construction of our coherent state $t=0$); as usual $t_0$ drops out from observables.

The time evolution operator $U$, connecting the interaction-picture operators to $\hat{\phi}$ and $\hat{\pi}$, is the usual one of interaction picture
\beq
U(t,t_0)&&=T\left\{e^{-i\int_{t_0}^tdt'H_I(t')} \right\}\\
&&=1+(-i)\int_{t_0}^tdt_1H_I(t_1)+(-i)^2\int_{t_0}^t dt_1\int_{t_0}^{t_1} dt_2 H_I(t_1)H_I(t_2)+\ldots\,.
\eeq
The last equality follows from the definition of the time-ordered product. We have chosen to expand it in terms of non-ordered products because the expectation value we are calculating is also non-ordered.

We evaluate the one-point expectation value \eqref{eq:perturbative_tadpole} to order $\lambda^2$ piece-by-piece, starting with the inner part
\beq
U(0,t) \FI (t,x) U(t,0)=\FI(t,x)+i\frac{\lambda}{3!}\int_0^t dt_1\int d^3z ~\FI^3(t_1,z)D(t_1-t,z-x)\nonumber \\
+\frac{\lambda^2}{3!4!}\int d^3z_1d^3z_2\Big\{-\int_0^tdt_1\int_{t_1}^tdt_2~\FI^4(t_1,z_1)\FI^3(t_2,z_2)D(t_2-t,z_2-x) \nonumber\\
+\int_0^tdt_1\int^{t_1}_0dt_2~\FI^3(t_1,z_1)\FI^4(t_2,z_2)D(t_1-t,z_1-x)\Big\}\nonumber\\
+\mathcal{O}(\lambda^3)\,.
\eeq
Here, the function $D$ stands for the following commutator
\beq
D(t_1-t,z-x)\equiv [\FI (t_1,z),\FI (t,x)]=\int \frac{d^3p}{(2\pi)^3}\frac{1}{2 E_p}\Big( e^{ip_\mu(z-x)^\mu}-e^{-ip_\mu(z-x)^\mu} \Big)\Big|_{p^0=E_p}\,.
\eeq

Moving forward, it is straightforward to show that
\beq
e^{i\phi_0\int d^3x' \hat{\pi}_{I}(0,x')}\FI(t,x)e^{-i\phi_0\int d^3x'' \hat{\pi}_{I}(0,x'')}=\Phi(t)+\FI(t,x)\,, \quad \text{with}\quad \Phi(t)\equiv \phi_0 {\rm cos}(mt)\,.
\eeq
Notice that the exponential operator still acts as a displacement operator but, unlike earlier calculations, instead of displacing the field by $\phi_0$ it does so by the time-dependent classical configuration of free theory. This is easy to understand, as field and momentum operators are at different times here, while in the earlier discussion they were at equal times; also, operators are the ones of the interaction picture, hence the appearance of the harmonic classical background. (Remember that earlier we saw that even in full theory one gets a full anharmonic classical background in similar calculation; the observation that was made perturbatively in time.)

Combining everything we have the following
\beq
A\equiv e^{i\phi_0\int d^3x' \hat{\pi}_{I}(0,x')}U(0,t) \FI (t,x) U(t,0)e^{-i\phi_0\int d^3x'' \hat{\pi}_{I}(0,x'')}-\Phi(t)=\FI(t,x)\nonumber \\
+i\frac{\lambda}{3!}\int_0^t dt_1\int d^3z ~(\Phi(t_1)+\FI(t_1,z))^3D(t_1-t,z-x)\nonumber \\
+\frac{\lambda^2}{3!4!}\int d^3z_1d^3z_2\Big\{-\int_0^tdt_1\int_{t_1}^tdt_2~(\Phi(t_1)+\FI(t_1,z_1))^4(\Phi(t_2)+\FI(t_2,z_2))^3D(t_2-t,z_2-x) \nonumber\\
+\int_0^tdt_1\int^{t_1}_0dt_2~(\Phi(t_1)+\FI(t_1,z_1))^3(\Phi(t_2)+\FI(t_2,z_2))^4D(t_1-t,z_1-x)\Big\}\nonumber\\
+\mathcal{O}(\lambda^3)\,.~~~~~~~~
\label{Ainter}
\eeq
Another quantity we need to calculate is $U(T,0)\hat{O}U(0,-T)$, with the inserted operator being the entire previous expression. Since we are interested in order $\lambda^2$ corrections and that the leading order term in the above expression is linear in the field (keeping in mind that the vacuum expectation value of the odd number of fields vanishes), we only require the following to the linear order in coupling
\beq
U(T,0)\hat{O}U(0,-T)=\hat{O}+(-i)\int_0^Td\tau~H_I(\tau)\hat{O}+(-i)\hat{O}\int_{-T}^0d\tau~H_I(\tau)+\mathcal{O}(\lambda^2)\,.
\eeq

Therefore, the entire one point function takes the following form
\beq
\la C| \hat{\phi}(t,x) |C\ra=\lim_{T\rightarrow \infty}\frac{1}{\la 0|U(T,-T) |0\ra}
\Big\{ \Phi(t)\la 0|U(T,-T)|0 \ra+\la 0|A|0\ra \nonumber~~~~~~~~~~~~~~~~~~~~~~~~~~~~\\
+\frac{\lambda^2}{3!4!}\int_0^tdt_1\int d^3zd^3y~D(t_1-t,z-x)\Big( \int_{-T}^Td\tau\la 0|\FI^4(\tau,y) |0\ra\Phi^3(t_1)
\nonumber\\
 +3\int_{0}^Td\tau \Phi(t_1)\la 0|\FI^4(\tau,y)\FI^2(t_1,z)|0\ra+3\int_{-T}^0d\tau \Phi(t_1)\la 0|\FI^2(t_1,z)\FI^4(\tau,y) |0\ra \Big)\Big\}\,;
\label{original}
\eeq
with $A$ given by \eqref{Ainter}. Notice that some of the disconnected contributions cancel trivially. One can easily recognize that the term with the integral over $\tau\in(-T,T)$ should cancels with the combination of normalisation pre-factor and one of the terms from $A$, just like disconnected diagrams cancel in the standard calculations.

Let us begin evaluating $\la 0|A|0\ra$ order by order in $\lambda$. To the zeroth order we obviously have simply $\Phi(t)$, which is the classical background of a free theory. The first order correction is given by
\beq
\la 0|A|0\ra=i\lambda\int_0^tdt_1\int d^3 z \left(\frac{1}{3!}\Phi^3(t_1)+\frac{1}{2}\Phi(t_1)\la \FI^2(t_1,z) \ra\right)D(t_1-t,z-x)+\mathcal{O}(\lambda^2)\,.~~~
\label{linearA}
\eeq
From this expression it immediately follows that, after substituting it into \eqref{original}, the order $\lambda$ part of $\la 0|U(T,-T) |0\ra$ in combination with the $\Phi^3$-term of \eqref{linearA} cancels with the first $\lambda^2$ term.
Notice that $\la \FI^2(t_1,z) \ra$ is an equal time and point expectation value and as such independent of the space-time coordinates. Henceforth, it will be denoted simply as $\la \FI^2 \ra$. After using the explicit expression for $D$, the result simplifies to
\beq
\la 0|A|0\ra=-\frac{\lambda \la \FI^2 \ra}{2}\cdot \phi_0\frac{t~{\rm sin}(mt)}{2m}+\frac{\lambda}{3!}\int_0^t dt_1 \Phi^3 (t_1) \frac{1}{m}{\rm sin}\left(m(t_1-t)\right)+\mathcal{O}(\lambda^2)\,.
\label{eq:propag}
\eeq
This result makes perfect physical sense. In fact, the first term together with $\Phi(t)$ in \eqref{original}, can be repackaged into $\phi_0\,{\rm cos}\left( \sqrt{m^2+\frac{\lambda}{2}\la \FI^2 \ra} ~t \right)$, up to $\lambda^2$ corrections, therefore corresponding to the mass renormalization. The third term, on the other hand, is simply the leading order classical correction due to nonlinearities, which can be easily integrated to give
\beq
\Phi^{(1)}(t)=-\frac{\lambda \phi_0^3}{3!}\frac{{\rm sin}(mt)(6mt+{\rm sin(2mt)})}{16m^2}\,.
\label{linsoln}
\eeq
Here, label '$(1)$' marks the fact that this correction is linear in coupling.

In order to compute quadratic corrections, we will need the correlation function
\beq
\la 0| \FI(t_1,z_1) \FI(t_2,z_2) |0\ra=\int \frac{d^3 p}{(2\pi)^3} \frac{1}{2 E_p} e^{i\vec{p}\cdot(\vec{z_1}-\vec{z_2})}\cdot e^{-iE_p(t_1-t_2)}\,;
\eeq
other correlators that make appearance in $A$-term and follows trivially from Wick's theorem are
\beq
\la 0| \FI(t_1,z_1) \FI^3(t_2,z_2) |0\ra=&&3 \la 0| \FI(t_1,z_1) \FI(t_2,z_2) |0\ra \cdot \la \FI^2 \ra\,,\\
\la 0| \FI^3(t_1,z_1) \FI ^3(t_2,z_2) |0\ra=&&6\la 0| \FI(t_1,z_1) \FI(t_2,z_2) |0\ra^3\nonumber \\
&&+3\la 0| \FI(t_1,z_1) \FI(t_2,z_2) |0\ra \cdot \la \FI^2 \ra^2,\\
\la 0| \FI^4(t_1,z_1) \FI ^2(t_2,z_2) |0\ra=&&12\la 0| \FI(t_1,z_1) \FI(t_2,z_2) |0\ra^2 \la \FI^2 \ra\nonumber \\
&&+3 \la \FI^2 \ra^3\,.
\eeq

Next we evaluate terms descending from the $\lambda^2$ part of \eqref{Ainter}, it is fairly straightforward to show that only terms with nontrivial power of $\la 0| \FI(t_1,z_1) \FI(t_2,z_2) |0\ra$ contribute,
\beq
\la 0|A|0\ra \Big|_{\lambda^2}=\Phi^{(2)}(t)+F_1(t)\la \FI^2 \ra+F_2(t)\la \FI^2 \ra^2+L_{coupling}+L_{sunrise}.
\label{appenA2}
\eeq
Here $\Phi^{(2)}(t)$ stands for the quadratic (in $\lambda$) corrections to the classical background and is given by
\beq
\Phi^{(2)}(t)=\frac{\lambda^2\phi_0^5}{4!^22^6m^4}\Big( (23-72m^2t^2){\rm cos}(mt)-24~{\rm cos}(3mt)+{\rm cos}(5mt)\nonumber \\
+12~mt(8~{\rm sin}(mt)-3~{\rm sin}(3 mt)) \Big)\,.
\label{quadsoln}
\eeq
The functions $F_{1,2}$ are given as follows
\beq
F_1(t)=-\frac{\lambda^2\phi_0^3}{4!2^5m^4}\Big( 2(-1+6m^2t^2){\rm cos}(mt)+2~{\rm cos}(3mt)+mt(-13~{\rm sin}(mt)+3~{\rm sin}(3mt)) \Big)\,,~\nonumber\\
F_2(t)=\frac{\lambda^2\phi_0}{32 m^3}t\Big( -mt~{\rm cos}(mt)+{\rm sin}(mt) \Big)\,.~~~~~~~~~~~~~~~~~~~~~~~~~~~~~~~~~~~~~~~~~~~~~~~~~~~~~~~~~~~~~\nonumber
\eeq
It is straightforward to check that \eqref{quadsoln} coincides with the $\lambda^2$-contribution to classical dynamics. Moreover, the second term of \eqref{appenA2} corresponds to 1-loop mass renormalization in  $\Phi^{(1)}(t)$. The third term of \eqref{appenA2}, on the other hand, contributes to 2-loop mass renormalization $\Phi(t)$. Unlike our previous results, here we get nontrivial loop diagrams as well. In particular, $L_{coupling}$ involves the loop-integral corresponding to the renormalization of coupling, $L_{sunrise}$ corresponds to the so called sunrise 2-loop diagram. To be more concrete, for the former we have
\beq
L_{coupling}=\frac{\lambda^2}{2}\int\frac{d^3p}{(2\pi)^3} \int_0^t dt_1\int_{0}^{t_1}dt_2~\frac{{\rm sin}(E_p(t_1-t_2))}{(2E_p)^2}\frac{{\rm sin}\left(m(t-t_1)\right)}{m}\nonumber \\
\times\left(\Phi(t_1)\Phi^2(t_2)+\Phi(t_1)\la \FI^2 \ra\right)\,.
\eeq
(Notice that terms containing $\la \FI^2 \ra$ correspond to two-loop diagram, in which one of the propagators in the loop has an attached bubble.) It is easy to check, that if we were to expand 1-loop part of $L_{coupling}$ in Taylor series in $t$, we would find that it is $\mathcal{O}(t^4)$. This seems to be consistent with the result of previous sections.

For the sunrise integral, on the other hand, we have
\beq
L_{sunrise}=\frac{\lambda^2}{3}\int\frac{d^3p}{(2\pi)^3}\frac{d^3p'}{(2\pi)^3}\frac{1}{2E_p\cdot 2E_{p'}\cdot 2E_{p+p'}}\times\nonumber\\
\times \int_0^t dt_1\int_{0}^{t_1}dt_2~{\rm sin}\Big((E_p+E_{p'}+E_{p+p'})(t_1-t_2)\Big)\frac{{\rm sin}\left(m(t-t_1)\right)}{m}\Phi(t_2)\,.
\eeq
Combining everything we have gotten so far and renormalising the mass we arrive at
\beq
\Phi(t)+\la 0|A|0\ra=&&\phi_0{\rm cos}(m_{\rm ph}t)-\frac{\lambda \phi_0^3}{3!}\frac{{\rm sin}(m_{\rm ph}t)(6m_{\rm ph}t+{\rm sin(2m_{\rm ph}t)})}{16m_{\rm ph}^2}\nonumber\\
&&+\frac{\lambda^2\phi_0^5}{4!^22^6m^4}\Big( (23-72m^2t^2){\rm cos}(mt)-24~{\rm cos}(3mt)+{\rm cos}(5mt)\nonumber \\
&&~~~~~~~~~~~+12~mt(8~{\rm sin}(mt)-3~{\rm sin}(3 mt)) \Big)\nonumber\\
&&+L_{coupling}+L_{sunrise}+\mathcal{O}(\lambda^3)\,,
\eeq
This already looks very promising, but before concluding let us take care of terms from \eqref{original} with $T$ as a limit of integration.

As a result of straightforward computation, all of the $T$-dependent terms cancel, except one; from which $T$ drops out in $T\rightarrow \infty$ limit. Putting everything together we have the following
\beq
\la C| \hat{\phi}(t,x) |C\ra=\Phi(t)+\la 0|A|0\ra- \frac{\lambda^2\la \FI^2 \ra}{2m}\int_0^tdt_1\int \frac{d^3p}{(2\pi)^3}\frac{1}{(2E_p)^3}\Phi(t_1){\rm sin}(m(t_1-t)){\rm cos}(2E_pt_1)\nonumber\\
+\mathcal{O}(\lambda^3)\,.~~~~~~~~~~~
\eeq
Obviously, the second term in this expression corresponds to a 2-loop correction, similar to $L_{sunrise}$, with 1-loop result nicely summarized in \eqref{finint}.

\end{document}